\documentclass[11pt,english]{article}
\usepackage[T1]{fontenc}
\usepackage[latin1]{inputenc}
\usepackage{a4wide}
\usepackage{subfigure}
\usepackage{amsmath}
\usepackage{graphicx}
\usepackage{amssymb}

\makeatletter
\usepackage{graphicx}
\makeatother \makeatletter \renewcommand\theequation{\hbox{\normalsize\arabic{section}.\arabic{equation}}} \@addtoreset{equation}{section} 

\renewcommand\thefigure{\hbox{\normalsize\arabic{section}.\arabic{figure}}} \@addtoreset{figure}{section} 

\renewcommand\thetable{\hbox{\normalsize\arabic{section}.\arabic{table}}} \@addtoreset{table}{section}

\makeatother 

\usepackage{babel}
\makeatother
\begin{document}
\date{October 15, 2003}

\title{\begin{flushright}{\normalsize ITP-Budapest Report No. 598}\end{flushright}\vspace{1cm}On
perturbative quantum field theory with boundary}

\author{Z. Bajnok$^{1}$%
\footnote{bajnok@afavant.elte.hu%
}, G. Böhm$^{2}$ and G. Takács$^{1}$%
\footnote{takacs@ludens.elte.hu%
}\\
$^{1}$\textit{\normalsize HAS Theoretical Physics Research Group,
Institute for Theoretical Physics, }\\
\textit{\normalsize Eötvös University, H-1117 Budapest Pázmány s.
1/A, Hungary}\\
\textit{\normalsize $^{2}$Research Institute for Particle Physics,
}\\
\textit{\normalsize H-1525 Budapest, 114 P.O.B. 49, Hungary}\\
\textit{\normalsize ~}\\
\textit{\normalsize ~}\\
\textbf{\textit{\normalsize This work is dedicated to Prof. Zalán
Horváth on the occasion of his 60th birthday.}}\\
\textbf{\textit{\normalsize ~}}\\
}

\maketitle
\begin{abstract}
Boundary quantum field theory is investigated in the Lagrangian framework.
Models are defined perturbatively around the Neumann boundary condition.
The analyticity properties of the Green functions are analyzed: Landau
equations, Cutkosky rules together with the Coleman-Norton interpretation
are derived. Illustrative examples as well as argument for the equivalence
with other perturbative expansions are presented. 
\end{abstract}

\section{Introduction}

In this paper we study quantum field theory with boundary in perturbation
theory. The boundary is a flat hypersurface with a space-like normal
vector, i.e. in appropriate inertial coordinates it is just given
by constraining one coordinate to take a constant value (conveniently
chosen to be zero). The motivation for such a study and most of the
explicit examples come from $1+1$ dimensional field theory, but we
present the formalism for an arbitrary number of spacetime dimensions
to show its generality.

Our main motivation comes from consideration of integrable boundary
QFT in 1+1 dimensions, although integrability is not required for
the general formalism to work. In such QFT, it is possible to construct
exact scattering amplitudes ($S$ matrices and reflection amplitudes)
using the bootstrap procedure. The central idea of the bootstrap is
that certain poles of scattering amplitudes give rise to new states
(bulk particles or excited boundary states) in the theory, which must
be treated as fundamental degrees of freedom in their own right ({}``nuclear
democracy''). 

However, in the general case there are poles which are not assigned
to any particle. In the case of bulk theories, bootstrap closure (or
consistency) means that all poles are explained in terms of intermediate
processes involving particles on-mass-shell, and that the residues
can be calculated using the on-shell couplings between the particles.
In case of poles associated to bound states, the intermediate state
is a one-particle state; the more general processes are associated
to so-called Coleman-Thun diagrams \cite{CT}. Consistency requires
that there exists a consistent choice for the on-shell couplings such
that the singular contributions at all the poles can be calculated
by the Cutkosky rules (in practice one normally looks only at the
most singular terms).

For QFT with boundaries, the respective rules for Coleman-Thun diagrams
were generalized by analogy in \cite{DTW,DM}, but there has been
no systematic derivation of these rules so far. In the bulk, the usual
derivation of singularity positions (Landau equations) and Cutkosky
rules uses the analytic properties of covariant perturbation theory
\cite{ELOP,IZ}, although the Cutkosky rules can also be understood
as a generalization of unitarity \cite{ELOP}.

There have been numerous previous works using perturbation theory
for boundary QFTs \cite{KIM1}--\cite{AbCo}. Unfortunately these
expansions are usually rather difficult to use to explore the analytic
structure, because the propagators do not have simple analytic properties
if expressed in terms of the momentum (in some cases even that is
impossible). Therefore in Section 2 we propose to use a perturbation
expansion around the free (Neumann) boundary condition and a constant
bulk field configuration. It is not at all evident that this approach
is correct in the general case, and the main difficulty comes from
the fact that the vacuum of the boundary QFT may be a non-constant
field configuration. As we mentioned above, expanding around such
a field configuration would make the investigation of the analytic
structure hopeless. Using a toy model, we present evidence in appendix
A that a re-summation of our perturbation expansion gives the correct
results that are expected from the standard approaches and in particular
the tadpole contributions reproduce the effect of the nontrivial background
field. As a further support for the correctness of the perturbative
expansion, in Appendix B we give a systematic derivation of one-loop
counter-terms in boundary sine-Gordon theory that have been used in
semiclassical calculations in the literature \cite{CD,CTa,KP}.

We also extend the perturbation theory by taking into consideration
fields living on the boundary, which is instrumental for the derivation
of the boundary Cutkosky rules. We define asymptotic states, and write
down the generalization of LSZ reduction formulae for boundary QFT,
which were previously derived in a restricted context by us \cite{BBT}. 

In Section 3, using the perturbative formulation, we derive the boundary
extension of Landau equations and Coleman-Norton interpretation, while
in Section 4 we obtain the boundary Cutkosky rules. In Section 5 we
present some explicit examples of singularities in the conjectured
scattering amplitudes of boundary sine-Gordon theory. The paper ends
by a discussion of the results and the outline of some issues that
remain to be investigated.

\section{Perturbation theory with a boundary}

In the course of analyzing of boundary quantum field theories we follow
the same line as presented in \cite{IZ} for bulk theories. Setting
the stage by introducing our conventions, we identify and canonically
quantize the free theory. The $R$ matrix is defined via asymptotic
states and is related to the Green functions using the boundary analogue
of the LSZ formula. Then we introduce Feynman rules for the computation
of Green functions, both in coordinate and momentum space.

\subsection{Conventions}

The coordinates describing the half space-time are\[
z=(t,\,\vec{x},\, y)\;,\;\vec{x}=\left(x^{1},\dots,x^{D-1}\right)\qquad,\qquad-\infty<t,\, x^{i}<\infty,-\infty<y\leq0\]
We denote the boundary coordinates by $x=(t,\vec{x})$ and use the
following abbreviations\begin{eqnarray*}
 &  & \int d\vec{x}=\prod_{i=1}^{D-1}\int_{-\infty}^{\infty}dx_{i}\quad,\quad\int dx=\int d\vec{x}\int_{-\infty}^{\infty}dt\quad,\quad\int dz=\int dx\int_{-\infty}^{0}dy\\
 &  & \vec{\nabla}=\left(\partial_{x^{1}},\dots,\partial_{x^{D-1}}\right)\end{eqnarray*}
Bulk fields are denoted by $\Phi_{\alpha}(z)$, while for boundary
fields we use $\phi_{a}(x)$. For simplicity, all the fields are supposed
to be real scalars (generalizing to other cases is straightforward).
The action is\begin{eqnarray}
S & = & \int dz\left\{ \frac{1}{2}\left[\left(\partial_{t}\Phi_{\alpha}\right)^{2}-\left(\vec{\nabla}\Phi_{\alpha}\right)^{2}-\left(\partial_{y}\Phi_{\alpha}\right)^{2}-M_{\alpha}^{2}\Phi_{\alpha}^{2}\right]-V\left(\Phi_{\alpha}\right)\right\} \nonumber \\
 & + & \int dx\left\{ \frac{1}{2}\left[\left(\partial_{t}\phi_{a}\right)^{2}-\left(\vec{\nabla}\phi_{a}\right)^{2}-m_{a}^{2}\phi_{a}^{2}\right]-U\left(\phi_{a},\Phi_{\alpha}\left(y=0\right)\right)\right\} \label{eq:generic_action}\end{eqnarray}
where \[
V\left(\Phi_{\alpha}\right)=\sum_{M\geq3}\sum_{\left\{ \alpha_{1},\dots\alpha_{M}\right\} }v_{\alpha_{1}\dots\alpha_{M}}\Phi_{\alpha_{1}}\dots\Phi_{\alpha_{M}}\]
 describes the bulk interaction, while\[
U\left(\Phi_{\alpha},\phi_{a}\right)=\sum_{M,N}\sum_{\left\{ \alpha_{1},\dots\alpha_{M}\right\} }\sum_{\left\{ a_{1},\dots a_{N}\right\} }u_{\alpha_{1}\dots\alpha_{M};a_{1}\dots a_{N}}\Phi_{\alpha_{1}}\dots\Phi_{\alpha_{M}}\phi_{a_{1}}\dots\phi_{a_{N}}\]
contains the bulk-boundary and pure boundary interaction terms (we
suppose that it contains no terms with $M=0$ and $N=1,2$).

\subsection{Free fields}

Here we give a short description of the free theory with $U=V=0$.
Free bulk fields satisfy the equations of motion\[
\left(\partial_{t}^{2}-\vec{\nabla}^{2}-\partial_{y}^{2}+M_{\alpha}^{2}\right)\Phi_{\alpha}(z)=0\quad,\quad\left.\partial_{y}\Phi_{\alpha}(z)\right|_{y=0}=0\]
The field satisfying the boundary condition can be decomposed as\begin{eqnarray*}
\Phi_{\alpha}\left(t,\vec{x},y\right) & = & \int_{-\infty}^{\infty}\frac{d\kappa}{2\pi}\cos(\kappa y)\int\frac{d\vec{k}}{\left(2\pi\right)^{D-1}}\exp(i\vec{k}\cdot\vec{x})\tilde{\Phi}_{\alpha}(\kappa,\vec{k},t)\\
 &  & \tilde{\Phi}_{\alpha}(\kappa,\vec{k},t)^{\dagger}=\tilde{\Phi}_{\alpha}(\kappa,-\vec{k},t)\quad,\quad\tilde{\Phi}_{\alpha}(\kappa,\vec{k},t)=\tilde{\Phi}_{\alpha}(-\kappa,\vec{k},t)\end{eqnarray*}
Similar decomposition is valid for the canonical momentum $\Pi_{\alpha}(z)=\partial_{t}\Phi(z)$.
The commutation relations are\[
\left[\Phi_{\alpha}(t,\vec{x},y),\Pi_{\beta}(t,\vec{x}',y')\right]=i\delta_{\alpha\beta}\delta(\vec{x}-\vec{x}')\left(\delta(y-y')+\delta(y+y')\right)\]
due to the Neumann boundary condition. We can introduce the bulk creation/annihilation
operators\begin{eqnarray*}
A_{\alpha}(\kappa,\vec{k},t) & = & i\tilde{\Pi}_{\alpha}(\kappa,\vec{k},t)+\Omega_{\alpha}(\kappa,\vec{k})\tilde{\Phi}_{\alpha}(\kappa,\vec{k},t)\\
A_{\alpha}(\kappa,\vec{k},t)^{\dagger} & = & -i\tilde{\Pi}_{\alpha}(\kappa,\vec{k},t)+\Omega_{\alpha}(\kappa,\vec{k})\tilde{\Phi}_{\alpha}(\kappa,\vec{k},t)\\
\Omega_{\alpha}(\kappa,\vec{k}) & = & \sqrt{\kappa^{2}+\vec{k}^{2}+M_{\alpha}^{2}}\end{eqnarray*}
that satisfy\[
\left[A_{\alpha}(\kappa,\vec{k},t),\, A_{\beta}(\kappa',\vec{k}',t)^{\dagger}\right]=\left(2\pi\right)^{D}2\Omega_{\alpha}(\kappa,\vec{k})\delta_{\alpha\beta}\delta(\vec{k}-\vec{k}')\left(\delta(\kappa-\kappa')+\delta(\kappa+\kappa')\right)\]
The boundary fields can be quantized in the usual way as free fields
living in $D$ dimensional spacetime:\begin{eqnarray*}
\phi_{a}(x) & = & \int\frac{d\vec{k}}{\left(2\pi\right)^{D-1}}\exp(i\vec{k}\cdot\vec{x})\tilde{\phi}_{a}(\vec{k},t)\quad,\quad\tilde{\phi}_{a}(\vec{k},t)^{\dagger}=\tilde{\phi}_{a}(-\vec{k},t)\\
\pi_{a}=\partial_{t}\phi_{a} & , & \left[\phi_{a}(t,\vec{x}),\pi_{b}(t,\vec{x}')\right]=i\delta_{ab}\delta(\vec{x}-\vec{x}')\\
a_{b}(\vec{k},t) & = & i\tilde{\pi}_{b}(\vec{k},t)+\omega_{b}(\vec{k})\tilde{\phi}_{b}(\vec{k},t)\quad,\quad\omega_{b}(\vec{k})=\sqrt{\vec{k}^{2}+m_{b}^{2}}\\
 &  & \left[a_{b}(\vec{k},t),\, a_{c}(\vec{k}',t)^{\dagger}\right]=\left(2\pi\right)^{D-1}2\omega_{b}(\vec{k})\delta_{bc}\delta(\vec{k}-\vec{k}')\end{eqnarray*}
The (normal ordered) free Hamiltonian can be written as \[
H=\frac{1}{2}\int_{-\infty}^{\infty}\frac{d\kappa}{2\pi}\int\frac{d\vec{k}}{\left(2\pi\right)^{D-1}}A_{\alpha}(\kappa,\vec{k},t)^{\dagger}A_{\alpha}(\kappa,\vec{k},t)+\int\frac{d\vec{k}}{\left(2\pi\right)^{D-1}}a_{b}(\vec{k},t)^{\dagger}a_{b}(\vec{k},t)\]
and the time development of the modes can be calculated as\[
A_{\alpha}(\kappa,\vec{k},t)=\mathrm{e}^{-i\Omega_{\alpha}(\kappa,\vec{k})t}A_{\alpha}(\kappa,\vec{k})\quad,\quad a_{b}(\vec{k},t)=\mathrm{e}^{-i\omega_{b}(\kappa,\vec{k})t}a_{b}(\vec{k})\]
The Fock space of free fields can be introduced in the standard way.
The vacuum satisfies\[
A_{\alpha}(\kappa,\vec{k})\left|0\right\rangle =0=a_{b}(\vec{k})\left|0\right\rangle \]
and the space is spanned by the states \[
A_{\alpha_{1}}(\kappa_{1},\vec{k}_{1})^{\dagger}\dots A_{\alpha_{M}}(\kappa_{M},\vec{k}_{M})^{\dagger}a_{b_{1}}(\vec{k}'_{1})^{\dagger}\dots a_{b_{N}}(\vec{k}'_{N})^{\dagger}\left|0\right\rangle \]
Due to the symmetry $A_{\alpha}(\kappa,\vec{k})=A_{\alpha}(-\kappa,\vec{k})$
we could constrain $\kappa\geq0$. However, it turns out to be simpler
to let $\kappa$ take general real values, and impose the former symmetry
property to account for the presence of the boundary.

The free propagators are\[
g_{ab}(x,x')=\left\langle 0\right|T\phi_{a}(x)\phi_{b}(x')\left|0\right\rangle =i\delta_{ab}\int\frac{dk}{\left(2\pi\right)^{D}}\frac{\mathrm{e}^{-ik\cdot(x-x')}}{k^{2}-m_{a}^{2}+i\varepsilon}\]
where $k=(k_{0},\vec{k})$, $\int dk=\int dk_{0}\int d\vec{k}$, and\begin{eqnarray*}
G_{\alpha\beta}(z,z') & = & \left\langle 0\right|T\Phi_{\alpha}(z)\Phi_{\beta}(z')\left|0\right\rangle \\
 & = & i\delta_{\alpha\beta}\int_{-\infty}^{\infty}\frac{d\kappa}{2\pi}\int\frac{dk}{(2\pi)^{D}}\frac{\mathrm{e}^{-ik\cdot(x-x')}}{k^{2}-\kappa^{2}-M_{\alpha}^{2}+i\varepsilon}\left(\mathrm{e}^{i\kappa(y-y')}+\mathrm{e}^{i\kappa(y+y')}\right)\end{eqnarray*}
The bulk propagator contains two terms: the first one describes the
direct propagation of the fields, while the other corresponds to reflection
on the free boundary. This can also be interpreted as a propagation
to the mirror image point at $-y'$, and will play an important role
in the formulation of the Feynman rules later.

\subsection{Asymptotic states, R matrix and reduction formulae}

We proceed with the usual assumptions: the interaction is switched
off adiabatically as $t\,\rightarrow\,\pm\infty$. The asymptotic
fields are then defined to be free fields $\Phi_{\alpha}^{as}$ and
$\phi_{a}^{as}$ ($as=in,\, out$), satisfying\begin{eqnarray*}
\lim_{t\rightarrow\mp\infty}\Phi_{\alpha}(t,\vec{x},y)-Z_{\alpha}\Phi_{\alpha}^{in/out}(t,\vec{x},y) & = & 0\\
\lim_{t\rightarrow\mp\infty}\phi_{a}(t,\vec{x})-Z_{a}\phi_{a}^{in/out}(t,\vec{x}) & = & 0\end{eqnarray*}
where the $Z$ coefficients, as usual, take care of the canonical
normalization of the fields, and the limits are understood in the
weak sense (i.e. for any matrix element of the fields). 

The modes of the asymptotic fields, which are obtained using the free
field mode expansions\begin{eqnarray*}
A_{\alpha}^{as}(\kappa,\vec{k}) & = & 2i\int_{-\infty}^{0}dy\int d\vec{x}\cos(\kappa y)\mathrm{e}^{i(\Omega_{\alpha}(\kappa,\vec{k})t-\vec{k}\cdot\vec{x})}\overleftrightarrow{\partial_{t}}\Phi_{\alpha}^{as}(t,\vec{x},y)\\
A_{\alpha}^{as}(\kappa,\vec{k})^{\dagger} & = & -2i\int_{-\infty}^{0}dy\int d\vec{x}\cos(\kappa y)\mathrm{e}^{-i(\Omega_{\alpha}(\kappa,\vec{k})t-\vec{k}\cdot\vec{x})}\overleftrightarrow{\partial_{t}}\Phi_{\alpha}^{as}(t,\vec{x},y)\\
a_{b}^{as}(\vec{k}) & = & i\int d\vec{x}\mathrm{e}^{i(\omega_{b}(\vec{k})t-\vec{k}\cdot\vec{x})}\overleftrightarrow{\partial_{t}}\phi_{b}^{as}(t,\vec{x})\\
a_{b}^{as}(\vec{k})^{\dagger} & = & -i\int d\vec{x}\mathrm{e}^{-i(\omega_{b}(\vec{k})t-\vec{k}\cdot\vec{x})}\overleftrightarrow{\partial_{t}}\phi_{b}^{as}(t,\vec{x})\end{eqnarray*}
create asymptotic states as follows:\begin{eqnarray*}
 &  & \left|\kappa_{1},\vec{k}_{1},\alpha_{1};\dots;\kappa_{M},\vec{k}_{M},\alpha_{M};\vec{k}_{1},b_{1};\dots;\vec{k}_{N},b_{N}\right\rangle _{as}=\\
 &  & A_{\alpha_{1}}^{as}(\kappa_{1},\vec{k_{1}})^{\dagger}\dots A_{\alpha_{M}}^{as}(\kappa_{M},\vec{k_{M}})^{\dagger}a_{b_{1}}^{as}(\vec{k_{1}})^{\dagger}\dots a_{b_{N}}^{as}(\vec{k_{N}})^{\dagger}\left|0\right\rangle \,.\end{eqnarray*}
We assume asymptotic completeness: both the $in$ and the $out$ states
form a complete basis. The unitary transformation between the two
is what we call the reflection matrix $R$ (in the usual terminology,
reflection matrix means the matrix element of $R$ between a bulk
one-particle $in$ and a bulk one-particle $out$ state). For any
given initial and final state, the corresponding matrix element of
the $R$ matrix (more precisely: $R$ operator) gives the probability
amplitude for the evolution of the initial state $\left|i\right\rangle $
at $t=-\infty$ into the final state $\left|f\right\rangle $ at $t=+\infty$:\[
R_{fi}=_{\quad out}\langle f\,|\, i\rangle_{in}\]
Now one can proceed to deduce reduction formulae for the matrix elements
between asymptotic states. The derivation is essentially the same
as in \cite{BBT}, so we only write down two examples of the resulting
formulae. Applying the formalism to an incoming bulk particle gives\begin{eqnarray*}
_{out}\left\langle A\right|\mathcal{O}|\kappa,\vec{k},\alpha;B\rangle_{in} & = & _{out}\left\langle A\right|\mathcal{O}A_{\alpha}^{in}(\kappa,\vec{k})^{\dagger}|B\rangle_{in}=\\
 & = & \mathrm{disconnected}\;\mathrm{part}\,+2iZ_{\alpha}^{-1/2}\int_{-\infty}^{0}dy\int dx\cos(\kappa y)\mathrm{e}^{-i(\Omega_{\alpha}(\kappa,\vec{k})t-\vec{k}\cdot\vec{x})}\times\\
 &  & \left\{ \partial_{t}^{2}-\partial_{y}^{2}-\vec{\nabla}^{2}+M_{\alpha}^{2}+\delta(y)\partial_{y}\right\} \,_{out}\left\langle A\right|T\mathcal{O}\Phi_{\alpha}(t,\vec{x},y)|B\rangle_{in}\end{eqnarray*}
where $\mathcal{O}$ denotes a general $T$ product of local operators.
The reduction formula for an incoming boundary particle created by
the $a$ modes are identical to the usual reduction formulae in $D$
space-time dimensions without boundary:\begin{eqnarray*}
_{out}\left\langle A\right|\mathcal{O}|\vec{k},b;B\rangle_{in} & = & _{out}\left\langle A\right|\mathcal{O}a_{b}^{in}(\vec{k})^{\dagger}|B\rangle_{in}=\\
 & = & \mathrm{disconnected}\;\mathrm{part}\,+iZ_{a}^{-1/2}\int dx\mathrm{e}^{-i(\omega_{\alpha}(\vec{k})t-\vec{k}\cdot\vec{x})}\times\\
 &  & \left\{ \partial_{t}^{2}-\vec{\nabla}^{2}+m_{a}^{2}\right\} \,_{out}\left\langle A\right|T\mathcal{O}\phi_{a}(t,\vec{x})|B\rangle_{in}\end{eqnarray*}
Formulae for outgoing particles can be written in a very similar form.

\subsection{Feynman rules in coordinate space}

In the interaction picture, the time evolution of the system can be
described by the operator\begin{eqnarray*}
U(t) & = & T\exp\left\{ -i\int_{-\infty}^{t}d\tau H_{int}\left(\tau\right)\right\} \\
H_{int}\left(\tau\right) & = & \int_{-\infty}^{0}dy\int d\vec{x}V\left(\Phi_{\alpha}^{in}(\tau,\vec{x},y)\right)+\int d\vec{x}U\left(\Phi_{\alpha}^{in}(\tau,\vec{x},y=0),\phi_{a}^{in}(t,\vec{x})\right)\end{eqnarray*}
The $R$ matrix can be expressed as\[
R=U(\infty)=T\exp\left\{ -i\int_{-\infty}^{\infty}d\tau H_{int}\left(\tau\right)\right\} \]
and the interacting fields take the form\begin{eqnarray*}
\Phi_{\alpha}(t,\vec{x},y) & = & U(t)^{-1}\Phi_{\alpha}^{in}(t,\vec{x},y)U(t)\\
\phi_{a}(t,\vec{x}) & = & U(t)^{-1}\phi_{a}^{in}(t,\vec{x})U(t)\end{eqnarray*}
Using the free field formulae and Wick's theorem, one can readily
derive the Feynman rules in coordinate space for the Green's functions
\begin{eqnarray*}
 &  & G_{\alpha_{1}\dots\alpha_{m};a_{1}\dots a_{n}}^{m,n}\left(\vec{x}_{1},y_{1},t_{1}\dots,\vec{x}_{m},y_{m},t_{m};\vec{x}_{1}',t_{1}',\dots\vec{x}_{n}',t_{n}'\right)=\\
 &  & \left\langle 0\right|T\Phi_{\alpha_{1}}(\vec{x}_{1},y_{1},t_{1})\dots\Phi_{\alpha_{m}}(\vec{x}_{m},y_{m},t_{m})\phi_{a_{1}}(\vec{x}_{1}',t_{1}')\dots\phi_{a_{n}}(\vec{x}_{n}',t_{n}')\left|0\right\rangle \end{eqnarray*}
from which the rules for the $R$ matrix can be obtained by applying
the reduction formulae. The resulting diagrams contain the following
ingredients:

\begin{enumerate}
\item Boundary propagator: \\
\\
\input{boundprop.pstex_t}
\item Direct bulk propagator\\
\\
\input{bulkdirect.pstex_t}
\item Reflected bulk propagator\\
\\
\input{reflectedbulk.pstex_t}
\item Bulk vertex (it doesn't make a difference whether a leg comes from
a direct or reflected bulk propagator)\\
\\
\input{bulkvertex.pstex_t}
\item Boundary vertex (once again, no distinction between the types of bulk
legs)\\
\\
\input{boundvertex.pstex_t}
\end{enumerate}
The vertex positions must be integrated over, for bulk vertices over
the bulk (half) space, while for boundary vertices over the boundary.
There are also combinatorial factors that follow straightforwardly
from counting the number of ways the particular Wick contractions
can be made. One has to take care to draw all possible diagrams obtained
by putting both the direct and the reflected propagators for each
bulk leg or line. In fact, for coordinate space rules we could have
used one type of bulk propagator as well: $G_{\alpha\beta}(z,z')$,
which consists of the sum of the two pieces. The separation of the
bulk propagator into two pieces, however, makes the momentum space
formulation much easier.

\subsection{Feynman rules in momentum space}

We can give the Feynman rules in momentum space by taking a simple
Fourier transform of the Green's functions:\begin{eqnarray*}
 &  & G_{\alpha_{1}\dots\alpha_{m};a_{1}\dots a_{n}}^{m,n}\left(\vec{k}_{1},\kappa_{1},\omega_{1}\dots,\vec{k}_{m},\kappa_{m},\omega_{m};\vec{k}_{1}',\omega_{1}',\dots\vec{k}_{n}',\omega_{n}'\right)=\\
 &  & \prod_{i=1}^{m}\left(\int_{-\infty}^{+\infty}dy_{i}\int dt_{i}d\vec{x}_{i}\mathrm{e}^{i(\omega_{i}t_{i}-\vec{k}_{i}\cdot\vec{x}_{i}-\kappa_{i}y_{i})}\right)\prod_{j=1}^{n}\left(\int dt_{j}d\vec{x}_{j}\mathrm{e}^{i(\omega_{j}'t_{j}'-\vec{k}_{j}'\cdot\vec{x}_{j}')}\right)\times\\
 &  & G_{\alpha_{1}\dots\alpha_{m};a_{1}\dots a_{n}}^{m,n}\left(\vec{x}_{1},y_{1},t_{1}\dots,\vec{x}_{m},y_{m},t_{m};\vec{x}_{1}',t_{1}',\dots\vec{x}_{n}',t_{n}'\right)\end{eqnarray*}
(the $y$ integrals are defined by a simple extension of the Green's
functions as even functions of $y$).

\begin{enumerate}
\item Boundary propagator: \\
\\
\input{mboundprop.pstex_t}
\item Direct bulk propagator\\
\\
\input{mbulkdirect.pstex_t}
\item Reflected bulk propagator\\
\\
\input{mreflectedbulk.pstex_t}
\item Bulk vertex\\
\input{mbulkvertex.pstex_t}\\
The momentum conservation for the $D$-momenta $k$ is the usual one,
but for the $\kappa$ component, the prime indicates that if two vertices
are connected by a reflected bulk propagator, the corresponding $\kappa$
must be oriented either outgoing or incoming at both vertices. For
direct bulk propagators, momentum is conserved as usual: all $D+1$
components of the momentum are oriented as outgoing at one end and
incoming at the other one. 
\item Boundary vertex\\
\input{mboundvertex.pstex_t}\\
There is no conservation law for the component $\kappa$, since the
position of boundary vertices is only integrated over along the boundary.
\end{enumerate}
There is also an integration over the momentum of each internal line.

As a result of the above rules, the presence of reflected bulk propagators
and boundary vertices breaks $\kappa$ conservation, as expected from
the absence of translational invariance in the direction perpendicular
to the boundary.

\section{Landau equations}

The analyticity properties of the Green functions can be analyzed
in the perturbative framework. The Landau equations describe the singularity
structure of the various Feynman diagrams which we are going to derive
for the boundary theory.

\subsection{Derivation of the Landau equations}

Consider a Feynman diagram in our generic scalar theory (\ref{eq:generic_action})
with $N$ outer bulk legs with momenta $(k_{i},\kappa_{i})\,,\; i=1,\dots,N$
and $M$ outer boundary legs with momenta $k_{i}\:,\; i=N+1,\dots,M+N$,
where $k_{i}$ are the components of the momenta parallel to the boundary
and $\kappa_{i}$ are the transverse components (only for bulk legs).
Using the fact that $D$-dimensional Poincare invariance (parallel
to the boundary) is unbroken, the resulting amplitude depends only
on the invariants $k_{i}\cdot k_{j}$ and $\kappa_{i}$. The general
Feynman integral can be written\begin{equation}
G=\int\prod_{i=1}^{L}\frac{d^{D}q_{i}}{(2\pi)^{D}}\prod_{j=1}^{K}\frac{d\chi_{j}}{2\pi}\prod_{r=1}^{I}(p_{r}^{2}-\pi_{r}^{2}-M_{r}^{2}+i\epsilon)^{-1}\prod_{s=I+1}^{I+J}(p_{s}^{2}-m_{s}^{2}+i\epsilon)^{-1}\label{eq:generic_Feynman}\end{equation}
where $q_{i}$ denote $D$-dimensional loop momenta for $L$ loops
in total, $\chi_{j}$ are the transverse {}``loop'' momenta (see
later), $(p_{r},\pi_{r})$ are the $D+1$-momenta of $I$ internal
bulk lines, while $p_{s}$ are the $D$-momenta of $J$ internal boundary
lines. Momentum conservation can be used to expressed $p$ and $\pi$
in terms of $q,\chi,k$ and $\kappa$ (taking into account that a
reflected bulk propagator reverses transverse momentum).

We can then introduce Feynman's parameterization in the usual way\begin{eqnarray*}
G & = & \int\prod_{i=1}^{L}\frac{d^{D}q_{i}}{(2\pi)^{D}}\prod_{j=1}^{K}\frac{d\chi_{j}}{2\pi}\prod_{r=1}^{I+J}\int_{0}^{1}d\alpha_{i}\delta\left(\sum\alpha_{i}-1\right)\psi(k,\kappa,q,\chi,\alpha)^{-I-J}\\
\psi(k,\kappa,q,\chi,\alpha) & = & \sum_{r=1}^{I}\alpha_{r}(p_{r}^{2}-\pi_{r}^{2}-M_{r}^{2}+i\epsilon)+\sum_{s=I+1}^{I+J}\alpha_{s}(p_{s}^{2}-m_{s}^{2}+i\epsilon)\end{eqnarray*}
and integrating out the loop momenta $q,\chi$ (these integrals are
-- apart from UV divergences taken care by counter-terms -- nonsingular
in general, just as in the case of bulk diagrams), we are left with
a multiple integral over the $\alpha$, which can be considered as
a function of the invariants $k_{i}\cdot k_{j}$ and $\kappa_{i}$.

Following the usual argument, a singularity can occur (in the limit
$\epsilon\,\rightarrow\,0$) when the hyper-contour in $\alpha$ space
is trapped between two singularities of the integrand, or a singularity
of the integrand takes place at the boundary of integration. The integrand
is singular if \begin{eqnarray}
 &  & \alpha_{r}=0\quad\textrm{or}\quad p_{r}^{2}-\pi_{r}^{2}-M_{r}^{2}=0\quad,\quad r=1,\dots,I\nonumber \\
 &  & \alpha_{s}=0\quad\textrm{or}\quad p_{s}^{2}-M_{s}^{2}=0\quad,\quad s=I+1,\dots,I+J\label{eq:Landau1}\end{eqnarray}
which are nothing other but straightforward generalizations of the
usual Landau equations. Reduced diagrams are defined by shrinking
internal lines with $\alpha_{i}=0$ to a point, and the remaining
lines must be on the mass-shell. A further requirement is that\begin{equation}
\frac{\partial}{\partial q_{i}}\left(\sum_{r=1}^{I}\alpha_{r}(p_{r}^{2}-\pi_{r}^{2}-M_{r}^{2}+i\epsilon)+\sum_{s=I+1}^{I+J}\alpha_{s}(p_{s}^{2}-m_{s}^{2}+i\epsilon)\right)=0\label{eq:Landau2}\end{equation}
and \begin{eqnarray}
\frac{\partial}{\partial\chi_{i}}\left(\sum_{r=1}^{I}\alpha_{r}(p_{r}^{2}-\pi_{r}^{2}-M_{r}^{2}+i\epsilon)+\sum_{s=I+1}^{I+J}\alpha_{s}(p_{s}^{2}-m_{s}^{2}+i\epsilon)\right)\nonumber \\
=-\frac{\partial}{\partial\chi_{i}}\left(\sum_{r=1}^{I}\alpha_{r}\pi_{r}^{2}\right) & = & 0\quad.\label{eq:Landau3}\end{eqnarray}
\begin{figure}
\begin{center}\includegraphics[%
  height=6cm,
  keepaspectratio]{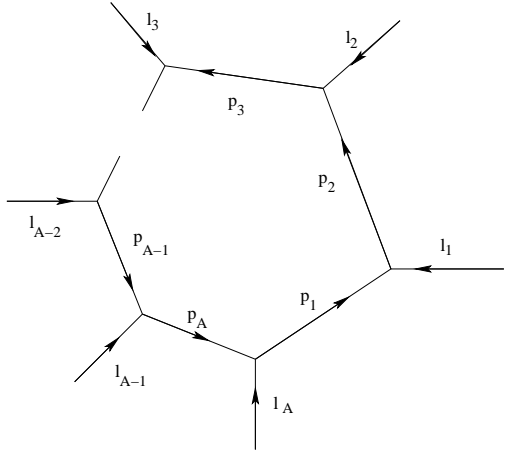}\end{center}

\caption{\label{cap:loop} Generic form of a closed loop}
\end{figure}
These equations can be brought into a more elegant form. First concentrate
on equation (\ref{eq:Landau2}) that is on the Lorentz invariant part.
Consider a closed loop in the Feynman diagram (Fig. \ref{cap:loop})
where the internal momenta are relabeled, the vertices can be of boundary
or of bulk type and the various $l$-s collect all the outer momenta
from the loop point of view, but in principle they can be outer $(k)$
or inner $(p)$ or their sums. Using momentum conservation the momentum
integration can be evaluated. As a result every $p_{i}\,,\, i=2\dots A$
can be expressed in terms of $p_{1}$ (which serves as the loop variable)
as follows\begin{equation}
p_{2}=p_{1}+l_{1}\;;\quad p_{3}=p_{2}+l_{2}=p_{1}+l_{1}+l_{2}\:;\:\dots\quad p_{A}=p_{1}+\sum_{j=1}^{A-1}l_{j}\label{eq:loopinp1}\end{equation}
and the overall momentum conservation $\sum_{i}l_{i}=0$ must hold.
Substituting (\ref{eq:loopinp1}) into (\ref{eq:Landau2}) for $q_{i}=p_{1}$
we obtain \begin{equation}
\sum_{\textrm{ each\, loop}}\alpha_{i}p_{i}=0\label{eq:landaubulk}\end{equation}
This argument is valid for a $D$ dimensional bulk theory as well
as for the $D$ non-transverse component of the momenta in a boundary
theory. 

\begin{figure}
\begin{center}\includegraphics[%
  height=6cm,
  keepaspectratio]{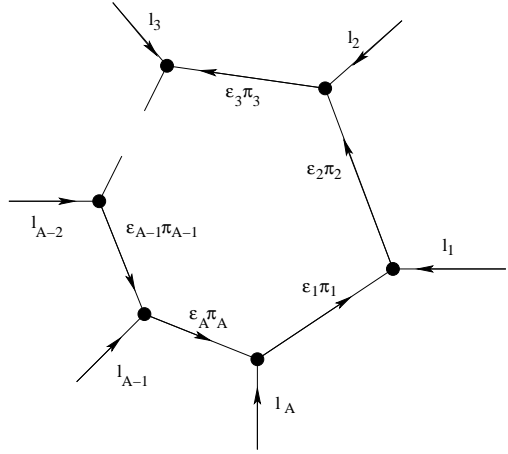}\end{center}

\caption{\label{cap:bloop} Generic form of a closed loop with bulk vertices
only}
\end{figure}

Now lets focus on the transverse momentum, that is on (\ref{eq:Landau3}).
Consider a loop as before but with bulk vertices only (Fig. \ref{cap:bloop}).
For bulk propagators we take $\epsilon=1$ while for the reflected
one $\epsilon=-1$. Using momentum conservation the momentum integration
can be eliminated successively giving rise to \[
\pi_{2}=\epsilon_{1}\pi_{1}+l_{1}\;;\quad\pi_{3}=\epsilon_{2}\pi_{2}+l_{2}=\epsilon_{1}\epsilon_{2}\pi_{1}+\epsilon_{2}l_{1}+l_{2}\:;\:\dots\quad\]
and in general\[
\pi_{i}=\prod_{j=1}^{i}\epsilon_{j}\pi_{1}+\prod_{j=2}^{i}\epsilon_{j}l_{1}+\dots+\prod_{j=k+1}^{i}\epsilon_{j}l_{k}+\dots+l_{i}\]
The delta function $\delta\left(\epsilon_{A}\pi_{A}+l_{A}-\pi_{1}\right)$
gives $\pi_{1}=\mu_{A}\pi_{1}+\dots$, where we have introduced $\mu_{i}=\prod_{j=1}^{i}\epsilon_{j}$
to encode the parity of the momentum change caused by reflected propagators.
Clearly if $\mu_{A}=-1$, that is the loop contains odd number of
reflected propagators the delta function eliminates the last integration
determining $\pi_{1}$ in terms of the $l$-s. In this case there
is no singularity. If, however, $\mu_{A}=1$ (loops with even number
of reflected propagator) $\pi_{1}$ is not determined, we have to
integrate over it ($\pi_{1}=\chi_{1})$ and we obtain the Landau equation:
\begin{equation}
\sum_{\textrm{ each\, loop}}\mu_{i}\alpha_{i}\pi_{i}=0\label{eq:landauloop}\end{equation}
\begin{figure}
\begin{center}\includegraphics[%
  height=5cm,
  keepaspectratio]{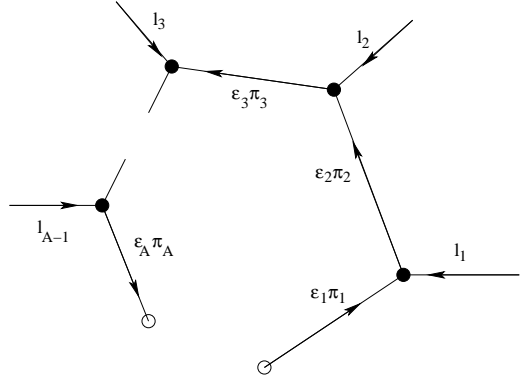}\end{center}

\caption{\label{cap:bouloop} Path starting from and ending at a boundary
vertex}
\end{figure}
Now consider a path in the Feynman graph starting and ending with
boundary vertices but containing bulk vertices only (Fig. \ref{cap:bouloop}).
Using the previous calculation $\pi_{A}$ can be expressed in terms
of $\pi_{1}$as before $\pi_{A}=\mu_{A-1}\pi_{1}+\dots$ but now there
is no more delta function so $\pi_{1}$ serves as a {}``loop'' variable,
$\chi_{1}$, for which integration has to be performed. As a result
we obtain the following Landau equation:\begin{equation}
\sum_{\textrm{ each\, path }}\mu_{i}\alpha_{i}\pi_{i}=0\label{eq:landaupath}\end{equation}
There is an interpretation of equation (\ref{eq:landaubulk}) in the
bulk theory in terms of electric network: it is just translated to
the usual Kirchoff laws for the $D$-momenta of the reduced diagram
with {}``resistances'' $\alpha_{r}$ and {}``currents'' $p_{r}$
so the {}``voltage drops'' are $\alpha_{r}p_{r}$. The generalization
for the boundary case: equation (\ref{eq:landauloop}) means that
$\alpha_{r}$-s are the {}``resistances'' and the $\pi_{r}$-s are
the {}``currents''. Note that there is one current for every component
of the $D+1$ dimensional energy-momentum vector and so for every
loop and vertex there are $D+1$ independent Kirchoff laws to write
down. 

The signs introduced by the reflected propagators can be traced as
follows: every reflected propagator carries an {}``inverter'' device
that changes the sign of the transverse momentum current, so $\mu_{r}$
keeps track of the actual sign of the current and $\mu_{r}\alpha_{r}\pi_{r}$
are the {}``voltage drops''. Equation (\ref{eq:landaupath}) means
that as far as the transverse momentum current is concerned, all the
boundary vertices are on an equipotential surface, which can be considered
as the reference point ({}``earth''). For the components of the
energy-momentum that are parallel to the boundary, the Kirchoff laws
take the usual form as in the bulk theory.

\subsection{Coleman-Norton interpretation}

The Landau equations for a generic Feynman graph in the bulk theory
are those we have in equation (\ref{eq:Landau1}) and (\ref{eq:landaubulk}).
Their physical region singularities have $\alpha_{i}\geq0$ and contain
real momenta $p_{i}$. Following Coleman and Norton \cite{CN}, these
equations can be interpreted as the existence of a space-time graph
of a process involving classical particles all on the mass shell,
all moving forward in time, and interacting only through energy and
momentum conserving interactions localized at space-time points. The
correspondence is one-to-one and goes as follows: for each internal
line draw a vector $\alpha_{r}p_{r}$, of length $\alpha_{r}M_{r}$.
(Lines with $\alpha_{r}=0$ are shrunk to a point). For each vertex
in the graph a space-time point is associated where the momentum conserving
interaction occurs. The consistency of this picture means, that two
different paths leading to the same vertex in the Feynman graph define
the same space-time point, which is nothing but the equation (\ref{eq:landaubulk}). 

\begin{figure}
\begin{center}\includegraphics[%
  height=5cm,
  keepaspectratio]{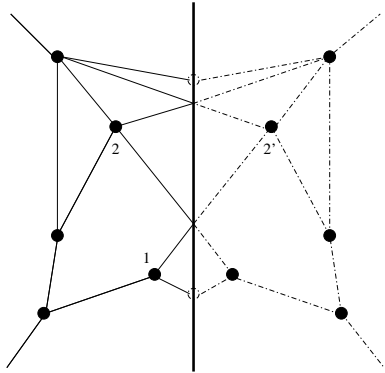}\end{center}

\caption{\label{cap:mirror} Space time diagram of a general on-shell process
using the method of images}
\end{figure}

The Coleman-Norton type interpretation of the boundary Landau equations
is the same as in the bulk case: the existence of a space-time graph
of a process involving particles all on the mass shell, all moving
forward in time, and interacting only through local interactions localized
at space-time points where now particles can scatter on the boundary,
which reverse the sign of the momenta. The correspondence goes as
follows: For each internal bulk line draw a vector $(\alpha_{r}p_{r},\alpha_{r}\pi_{r})$
with length $\alpha_{r}M_{r}$ (reflected line in the case of a reflected
propagator), while for boundary lines a vector $(\alpha_{s}p_{s})$
of length $\alpha_{s}m_{s}$ (lying in the boundary). Bulk vertices
are located at space-time bulk points where the momentum preserving
interaction occurs. Boundary vertices are located at the boundary
points with interactions where the transverse component of the momentum
is not conserved. To see that for a space-time diagram the Landau
equations are satisfied consider the diagram in Fig. \ref{cap:mirror}.

On the left side of the boundary is the real space-time picture of
some process, while on the right hand side is its mirror image with
respect to the boundary, which is necessary to introduce in order
to interpret it in terms of the Landau equations. Instead of drawing
a reflected line for a reflected propagator (e.g. from $1$ to $2$)
we consider a straight line from the real world to the mirror one
(from $1$ to $2'$) or vice versa. The variable $\mu$ keeps track
on which side of the boundary we are and to which {}``world'' (real
or mirror image) the next line is directed. Consistency of the Coleman-Norton
picture means that for two different paths between two interaction
point the displacement vectors must sum up to the same value, which
is ensured by equation (\ref{eq:landauloop}). Evidently we compare
only points that are both on the {}``real'' side of the boundary
so it is sufficient to consider loops with even number of reflected
propagators. Another consistency condition is that all the boundary
vertices must have the same transverse coordinate, which translates
into equation (\ref{eq:landaupath}). 

In order to reverse the above argument, we must show that to any solution
of the Landau equations, a space-time diagram can be associated. This
can be done by drawing the extended diagram first, similarly as we
would draw in a bulk theory, then choosing a loop with odd number
of reflected propagator and determining the location of the boundary
by identifying the endpoints of the closed loop as the mirror image
of each other. Clearly such a loop is not closed in the transverse
coordinate. Now the Landau equations for the loops with even number
of reflected propagators ensure that the location of the boundary
does not depend on the choice of the loop with odd number of reflected
propagators.

\section{Cutkosky rules}

\subsection{Derivation}

A further important ingredient is provided by the boundary analogue
of Cutkosky rules to calculate discontinuities across the cuts in
the physical region determined by the solutions of the Landau equations.
To give the boundary generalization of Cutkosky rules we need a further
definition. Note that if we write down a Feynman integral $G$ of
the form (\ref{eq:generic_Feynman}) with $I$ bulk internal lines,
each bulk line corresponds to either a direct or a reflected propagator
so there are actually $2^{I}$ distinct diagrams that differ only
which bulk lines are taken to be direct or reflected (represented
by continuous resp. dashed lines in the associated Feynman diagram).
Let $C(G)$ denote the sum of terms the possible assignments of direct/reflected
bulk lines. Each individual term can be written in the form (\ref{eq:generic_Feynman}),
the difference being the expression of the loop momenta $\chi$ in
terms of the internal momenta $\kappa$ due to the different conservation
laws obeyed by the two different types of bulk lines.

It is clear that $C(G)$ can have a singularity at a given value of
external momenta only if some individual term in the sum does. For
an integral of the form (\ref{eq:generic_Feynman}), every singularity
corresponds to a solution of the Landau equations (\ref{eq:Landau1},\ref{eq:Landau2},\ref{eq:Landau3}).
In such a solution there are internal lines which have a corresponding
$\alpha_{r}$ that is nonzero (and so are on the mass shell); these
are the ones that make up the reduced diagram corresponding to the
given singularity. We remark that since the momentum conservation
conditions (\ref{eq:Landau3}) are different for each individual term
in the sum $C(G)$, in general only a few terms give a nonzero contribution
to a given singularity.

Let us denote the set of lines which are on-mass-shell at the singularity
by $G_{0}$. We restrict ourselves to cases when the graph $G\setminus G_{0}$,
obtained by severing the lines on-mass-shell separates into two disconnected
graphs $G_{1}$ and $G_{2}$ (in the bulk the corresponding singularities
are called {}``normal thresholds'').

Then the rules are that the discontinuity of $C(G)$ (\ref{eq:generic_Feynman})
across the cut corresponding to the singularity can be evaluated by
substituting the propagators of the internal lines for which $\alpha_{r}\neq0$
as follows:\begin{eqnarray*}
\frac{1}{p_{r}^{2}-\pi_{r}^{2}-M_{r}^{2}+i\epsilon} & \rightarrow & -2\pi i\delta^{(+)}\left(p_{r}^{2}-\pi_{r}^{2}-M_{r}^{2}\right)\\
\frac{1}{p_{r}^{2}-m_{r}^{2}+i\epsilon} & \rightarrow & -2\pi i\delta^{(+)}\left(p_{r}^{2}-m_{r}^{2}\right)\end{eqnarray*}
change $\epsilon\rightarrow-\epsilon$ for internal lines in one of
the components (say $G_{2}$, the other choice just gives the jump
in the opposite direction across the cut), and then perform the integration
in (\ref{eq:generic_Feynman}). Here $\delta^{(+)}$ denotes a delta
function where only the root $p_{0}=+\sqrt{\vec{p}_{r}^{2}+\pi_{r}^{2}+M_{r}^{2}}$
or $+\sqrt{\vec{p}_{r}^{2}+m_{r}^{2}}$ is taken into account.

To show this, we generalize a very elegant proof for the bulk Cutkosky
rules, originally due to Nakanishi. The bulk proof is spelled out
in detail in \cite{IZ}, so we only give the details necessary for
its extension to the boundary case. We construct an auxiliary field
theory to any given diagram class $C(G)$ in the following way. To
each internal bulk line of the diagram we associate a different species
of bulk field $\Phi_{r}$ of mass $M_{r}$ ($r=1,\dots,I$), similarly
to each internal boundary line a boundary field $\phi_{s}$ of mass
$m_{s}$ ($s=1,\dots,J$). To every bulk vertex $v$ with some external
bulk line we attach a field $\tilde{\Phi}_{v}$ of mass $\tilde{M}_{v}$,
where $\tilde{M}_{v}^{2}=P_{v}^{2}$, $P_{v}$ is the total $(D+1)$-momentum
entering the given vertex from its external legs. Similarly to every
boundary vertex $u$ we attach a boundary field $\tilde{\phi}_{u}$
of appropriate mass $\tilde{m}_{u}$, $\tilde{m}_{u}^{2}=p_{u}^{2}$
where $p_{u}$ is the total $D$-momentum entering the vertex from
external bulk and boundary lines. The auxiliary action corresponding
to $C(G)$ is defined as \begin{eqnarray*}
\mathcal{A_{\mathrm{\mathcal{\mathsf{G}}}}} & = & \int dz\sum_{r=1}^{I}\frac{1}{2}\left[\left(\partial\Phi_{r}\right)^{2}-M_{r}^{2}\Phi_{r}^{2}\right]+\sum_{v}\strut^{\prime}\frac{1}{2}\left[\left(\partial\tilde{\Phi}_{v}\right)^{2}-\tilde{M}_{v}^{2}\tilde{\Phi}_{v}^{2}\right]+\sum_{v}\tilde{\Phi}_{v}\prod_{r\rightarrow v}\Phi_{r}+\\
 &  & \int dx\sum_{s=1}^{J}\frac{1}{2}\left[\left(\partial\phi_{s}\right)^{2}-m_{s}^{2}\phi_{s}^{2}\right]+\sum_{u}\strut^{\prime}\frac{1}{2}\left[\left(\partial\tilde{\phi}_{u}\right)^{2}-\tilde{m}_{u}^{2}\tilde{\phi}_{u}^{2}\right]+\sum_{u}\tilde{\phi}_{u}\prod_{r\rightarrow u}\Phi_{r}\prod_{s\rightarrow u}\phi_{s}\end{eqnarray*}
where $\sum_{v}'$ and $\sum_{u}'$ mean that the sum goes only for
bulk/boundary vertices that have external lines entering them, $\tilde{\Phi}_{v}$
and $\tilde{\phi}_{u}$ must be put equal to $1$ for vertices which
connect only to internal lines, and expressions of the type $r\rightarrow u$
means taking product over internal lines $r$ that connect to vertex
$u$.

The crucial point is that $C(G)$ gives the lowest order contribution
to the scattering amplitude with the given external momenta in the
auxiliary theory. Let $i$ denote the initial state of this scattering
and $f$ the final one (external lines attached to $G_{1}$ and $G_{2}$,
respectively). Then the transition amplitude for this process can
be written\[
\mathcal{T}_{fi}=C(G)\]
However, on-shell processes are unitary, thus we have \[
\mathcal{T}_{fi}-\mathcal{T}_{if}^{*}=\sum_{G_{0}}C\left(G_{1}\right)C\left(G_{2}\right)^{*}\]
where $\sum_{G_{0}}$ means a phase space summation over the lines
involved in $G_{0}$. Following a similar reasoning as in \cite{IZ}
yields exactly the rules spelled out above.

\subsection{Comments on the general case}

For different topologies, we expect that similar rules hold. The reason
is simple: the Cutkosky rules themselves depend only on the analytic
structure of propagators. In the perturbation theory around the free
(Neumann) boundary conditions we see that the momentum space form
of propagators is just the same as in the bulk theory, the only difference
comes from the momentum conservation rules which express the internal
momenta in terms of the external and loop momenta and from the fact
that boundary propagators have vanishing transverse momentum. However,
insofar as these expressions are linear (which is true even in the
boundary situation), neither the precise form of the conservation
rules enters in any derivation of the Cutkosky rules, nor the constraint
on the transverse momentum for boundary lines (which is also linear)
is relevant. Therefore we can see that the boundary Cutkosky rules
given above can be derived for any topology for which a derivation
is given in the bulk (e.g. the well-known triangle anomalous threshold
diagram, which is not in the class covered by Nakanishi's derivation). 

In the above derivation we also supposed that $\tilde{M}_{v}^{2}$
and $\tilde{m}_{u}^{2}$ are positive, which corresponds to singularities
in the physical region. Beyond the physical region nontrivial analytic
continuation is required similarly to the bulk case. Extension of
Cutkosky rules to such singularities is a very complicated issue,
tantamount to a generalization of unitarity. In this paper we do not
enter into further discussion of singularities in the non-physical
region, but simply suppose that an extension of these rules can be
worked out.

Finally we note that singular contributions corresponding to the same
reduced diagram can be summed up, giving the familiar result that
the total contribution (to all orders of perturbation theory) can
be obtained by substituting the exact vertex functions for the vertices
in the reduced diagram characterizing the singularity (see e.g. Section
2.9. of \cite{ELOP}). 

One has to be very careful in deducing the singularities of the Green
functions from the analyzis of the perturbative series. Sometimes
individual terms may have singularities that are cancelled when a
sum of terms is taken. Some other times individual terms do not yield
the correct singularities at all. It happens when some nonperturbative
dynamical property, such as formation of a bound state, is involved.
To handle these situations a complete or a partial summation of the
perturbative series has to be performed.

\subsection{Infrared divergences}

As shown in appendix A.2, perturbation theory around the Neumann boundary
condition has infrared divergences. These arise from on-shell bulk
particles propagating along the boundary as expected from the Landau
equations and correspond to a boundary bound state. 

However, if there is a quadratic term of the form $\lambda\Phi(y=0)^{2}/2$
for the corresponding bulk field $\Phi$ in the boundary potential
$V$ (which is the case in general), then the infrared divergence
is nonphysical. There are infinitely many infrared divergent terms
with different degree of singularity and resummation in the coupling
$\lambda$ shifts the singularity from $k=0$ to $k=-i\lambda$, and
it does not affect the derivation of the physical region Cutkosky
rules presented above. 

For $\lambda>0$ the singularity is shifted to the unphysical sheet
of the scattering amplitude, so it is physically irrelevant. For $-m<\lambda<0$,
the pole is again shifted away from the region of physical values
of the incoming momentum, however, it remains on the physical sheet
and appears as a boundary bound state (see A.2). Note that the derivation
of the Cutkosky rules given above is perturbative, but as it rests
on general principles such as analiticity and unitarity, we expect
the rules to extend to singularities corresponding to such nonperturbative
intermediate states as well, just as in bulk theories. In fact, the
applications presented in the next Section require such an extension
of Cutkosky rules to singularities with complex values of incoming
momenta, but still in the physical sheet.

Finally, let us remark that the derivation of the Landau equations
and the Nakanishi proof for the Cutkosky rules for physical region
singularities can be performed in a perturbation theory that starts
around a nonzero value of $\lambda$. In this case the reflected bulk
propagator must be changed to\[
\frac{i}{k_{r}^{2}-\kappa_{r}^{2}-M_{r}^{2}+i\epsilon}\;\frac{\kappa_{r}-i \lambda}{\kappa_{r}+i\lambda}\]
For amplitudes in the physical region, the last factor is just a nonsingular
(phase-valued) expression, and so does not affect the derivations
in any essential way.

\section{Applications}

In the following we restrict ourselves to $1+1$ spacetime dimensions,
since the example theory we use is sine-Gordon theory with integrable
boundary conditions. The above formalism is of course more general,
however, integrable field theories make possible an exact and non-perturbative
verification of the general scheme.

\subsection{Boundary bound state poles}

\begin{figure}
\begin{center}\includegraphics{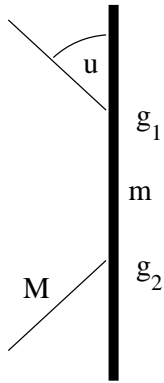}\end{center}

\caption{\label{cap:spole} Diagram corresponding to an on-shell boundary
state}
\end{figure}
The simplest application of the Cutkosky rules is when there is a
single intermediate line on-shell. If that is a bulk line, then we
get back the well-known result for a first order pole corresponding
to an intermediate particle in the bulk. If the line in question is
a boundary line, we get a discontinuity of the form\[
2\pi g_{1}g_{2}^{*}\delta^{(+)}\left(E^{2}-m^{2}\right)\]
corresponding to the diagram Fig. \ref{cap:spole}, where $E=M\cosh\vartheta$
and $m=M\cos u$ for a pole at imaginary rapidity $\vartheta=iu$. 

This reproduces the result of Ghoshal and Zamolodchikov \cite{GZ},
which states that an intermediate on-shell boundary state produces
a pole in the reflection factor $R\left(\vartheta\right)$ of the
form\[
R\left(\vartheta\right)\sim\frac{1}{2}\frac{if_{1}f_{2}^{*}}{\vartheta-iu}\]
where the dimensionless couplings $f_{1,2}$ are related to $g_{1,2}$
by some normalization factors, coming partially from converting the
$\delta^{(+)}$ to a function of $\vartheta$ and also from relating
the reflection factor in terms of the Feynman amplitude, but the precise
form of this relation is not important for us here.

\subsection{Coleman-Thun diagrams}

We present some examples for more complicated intermediate diagrams
following Coleman and Thun \cite{CT} (see also \cite{DTW,DM,BPT}
for the boundary case). To be specific, we consider sine-Gordon theory
with integrable boundary conditions. The groundwork for determining
its spectrum and scattering amplitudes was laid down in \cite{GZ,G},
while the bootstrap closure of this theory for general boundary conditions
was obtained in \cite{BPT}, following the work in \cite{DM}.

Before proceeding to the examples, we introduce a useful notation.
Let $f(\vartheta)$ be a meromorphic function of one complex variable
$\vartheta$. Let us suppose that the Laurent expansion of $f$ around
$\vartheta=\vartheta_{0}$ takes the form \[
f(\vartheta)=A\left(\vartheta-\vartheta_{0}\right)^{n}+\sum_{k>n} b_{k}\left(\vartheta-\vartheta_{0}\right)^{k}\qquad,\qquad A\neq0\;.\]
We then define\[
f\left[\vartheta_{0}\right]=A.\]
If $\vartheta_{0}$ is not a pole or a zero of $f$, then this is
just the value of the function. If $\vartheta_{0}$ is a first order
pole, then this is just the residue, while for a first order zero
it is the derivative of $f$ at $\vartheta_{0}$.

\begin{figure}
\begin{center}\includegraphics{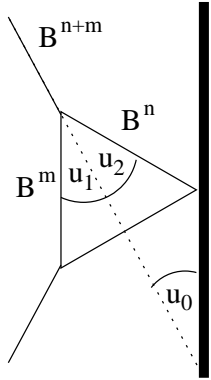}\end{center}

\caption{\label{cap:triangle} Coleman-Thun graph for a boundary-independent
pole in the breather reflection factor. The angles are $u_{0}=u_{1}=n\pi/2\lambda$
and $u_{2}=m\pi/2\lambda$.}
\end{figure}

The first family we consider are first order poles independent of
the boundary conditions and occur in reflection factors of breathers
on any boundary state (ground state and excited states alike). Let
us denote the reflection factor of a breather $B^{k}$ with rapidity
$\vartheta$ by $R^{(k)}\left(\vartheta\right)$ (we omit the specification
of the boundary state, as the argument does not depend on it). Then
$R^{(k)}$ has a pole at $\vartheta=in\pi/2\lambda\;,\; n=1,\dots,k-1$.
We pick a pole with a given $n$ and write $k=m+n$. Following \cite{BPT}
this pole can be associated with the scattering process in diagram
\ref{cap:triangle} (we recall that an imaginary rapidity difference
$\vartheta=iu$ corresponds to a Euclidean angle $u$ in the plane).
The diagram obviously gives a first-order pole, since it has $3$
propagators (one $\delta^{(+)}$ factors each), and a loop integral
which means $2$ integrations, leaving us with a single $\delta$
function in the discontinuity which is characteristic of a first order
pole. 

\begin{figure}
\begin{center}\subfigure[$u\neq u_0$]{\includegraphics{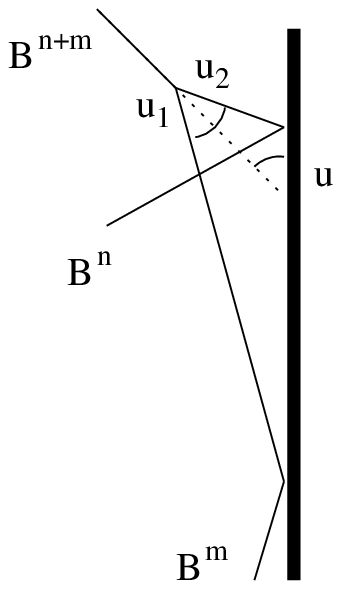}}\subfigure[$u\,\rightarrow\, u_0$ limit]{\includegraphics{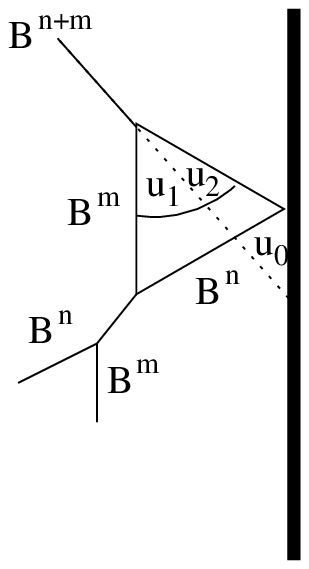}}\end{center}

\caption{\label{cap:triang_calc} Calculating the contribution of the triangle
graph using the bootstrap}
\end{figure}

To establish that this process explains the pole of the reflection
factor, we need to calculate its contribution to the residue. The
straightforward method would be to write down the Cutkosky rules directly
and calculate the integrals, which do give a contribution of the right
form. However, the exact coefficient is not easy to obtain as one
needs to ensure the right normalization of the reflection amplitude
first, and then also extract the breather-breather fusion coupling
$g_{nm}^{n+m}$ from the $S$ matrix, once again taking care of all
details of normalization.

Instead, we can follow a simpler route, noting that the $S$ matrices
and the reflection amplitudes satisfy the bootstrap which entails
a vast number of identities. We can tune the incoming rapidity away
from $\vartheta=iu_{0}$, obtaining the process in Fig. \ref{cap:triang_calc}
(a), which, by virtue of factorized scattering, has the amplitude\[
g_{nm}^{n+m}R^{(n)}\left(\vartheta+iu_{1}\right)R^{(m)}\left(\vartheta-iu_{2}\right)S\left(2\vartheta+i\left(u_{1}-u_{2}\right)\right)\qquad,\qquad\vartheta=iu\]
Taking $\vartheta\,\rightarrow\, iu_{2}$ we hit a pole in the $S$
matrix. The contribution of this pole reads\[
g_{nm}^{n+m}R^{(n)}\left(i(u_{2}+u_{1})\right)(-1)\frac{1}{2}S\left[i\left(u_{1}+u_{2}\right)\right]\]
On the other hand, in the same limit the diagram in Fig. \ref{cap:triang_calc}
(a) becomes the one drawn in Fig. \ref{cap:triang_calc} (b), from
which amplitude of the triangle graph Fig. \ref{cap:triangle} can
be obtained by dividing with the breather fusion coupling $g_{nm}^{n+m}$.
This means that to explain the pole in the reflection factor of $B^{m+n}$,
the following identity must hold:\[
R^{(m+n)}\left[iu_{2}\right]=-\frac{1}{2}R^{(n)}\left(i(u_{2}+u_{1})\right)S\left[i\left(u_{1}+u_{2}\right)\right]\]
which can be verified by direct substitution.

\begin{figure}
\begin{center}\includegraphics{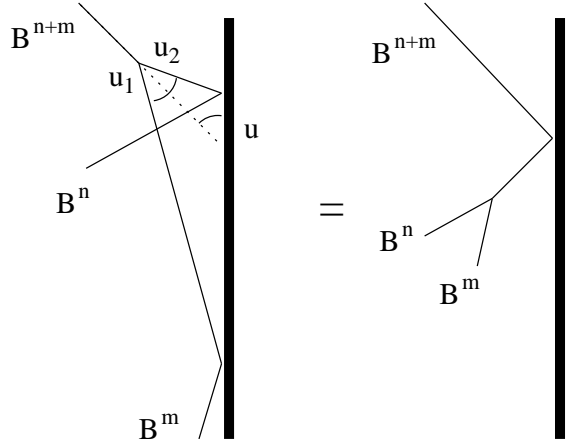}\end{center}

\caption{\label{cap:triang_bootstrap} Bootstrap relation for $R^{(m+n)}$}
\end{figure}

It is very interesting to note that this identity is guaranteed by
the bootstrap construction of the reflection factor $R^{(m+n)}$,
using the relation drawn in Fig. \ref{cap:triang_bootstrap}.

\begin{figure}
\begin{center}\subfigure[Breather direct]{\includegraphics[%
  height=6cm,
  keepaspectratio]{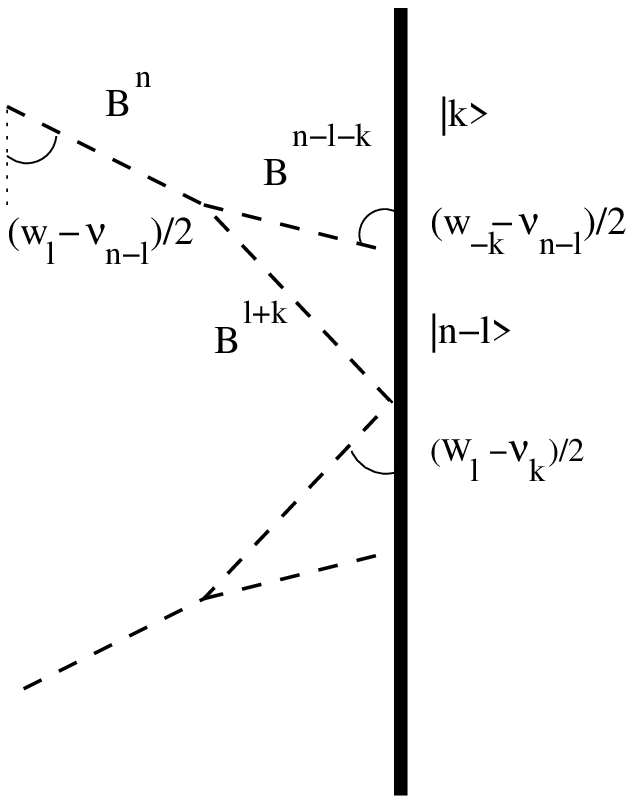}}~~~~~~~~~~\subfigure[Breather decay]{\includegraphics[%
  height=6cm,
  keepaspectratio]{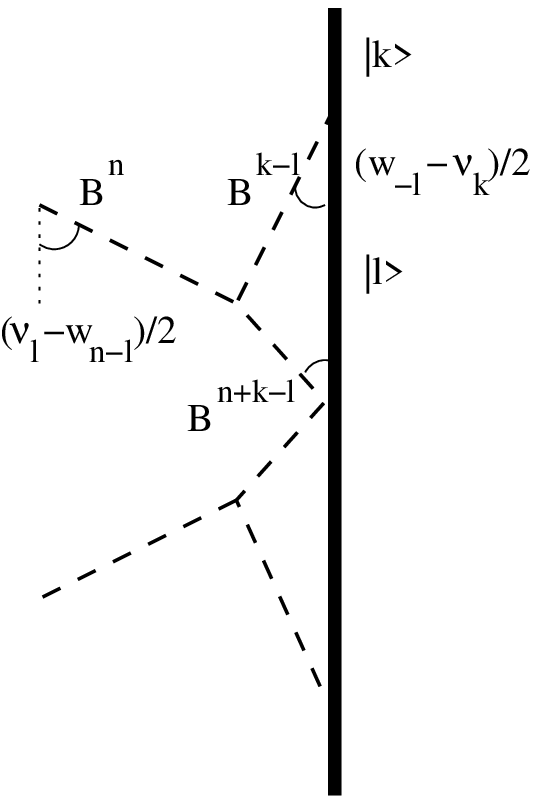}}\end{center}

\caption{\label{cap: bdep} Coleman-Thun explanation for boundary dependent
poles in breather reflection factors}
\end{figure}

We have also considered two other classes of Coleman-Thun diagrams,
using similar arguments. Fig. \ref{cap: bdep} (a) shows a process
where an incoming breather splits into two breathers of lower index,
one of which excites the boundary state on which the other gets reflected.
The process in Fig. \ref{cap: bdep} (b) describes an incoming boundary
state decaying by the emission of a breather that fuses with the incoming
breather to a higher one, which in turn gets reflected from the boundary.
The notations for the boundary states $|n\rangle$ and the fusing
angles $\nu_{k},w_{l}$ are explained in \cite{BBT}.

Naively, these diagrams would give a second order pole since the number
of propagators is $6$ (including propagators for boundary states
as dictated by the Cutkosky rules), while there are two loop integrals
which give the order $6-2\times2=2$. However, the reflection at the
middle of these diagrams takes place at a rapidity at which the corresponding
amplitude has a first-order zero, which results in a first-order contribution.

Straightforward application of bootstrap principles shows that the
contribution of these diagrams to the given residues can be written
as\begin{eqnarray*}
(a): & \quad & -R_{|k\rangle}^{(n-l-k)}\left[\left(w_{-k}-\nu_{n-l}\right)/2\right]R_{|n-l\rangle}^{(l+k)}\left[\left(w_{l}-\nu_{k}\right)/2\right]S^{n-l-k,l+k}\left[n\pi/2\lambda\right]\\
(b): & \quad & -R_{|k\rangle}^{(k-l)}\left[\left(w_{-l}-\nu_{k}\right)/2\right]R_{|l\rangle}^{(n+k-l)}\left[\left(\nu_{k-n}-w_{-l}\right)/2\right]S^{n,k-l}\left[(n+k-l)\pi/2\lambda\right]\end{eqnarray*}
If these diagrams are to explain the residues, then the above expressions
must be equal to\begin{eqnarray*}
(a): & \quad & R_{|k\rangle}^{(n)}\left[\left(w_{l}-\nu_{n-l}\right)/2\right]\\
(b): & \quad & R_{|k\rangle}^{(n)}\left[\left(\nu_{l}-w_{n-l}\right)/2\right]\end{eqnarray*}
We checked these equalities by direct evaluation. As before, they
are also a consequence of the bootstrap relations between the reflection
factors and $S$ matrices.

It is quite remarkable that the identities necessary for the diagrams
to explain the residues of the pole hold by the bootstrap. This seems
to be the case for breather reflection factors in sine-Gordon theory
for generic values of the boundary parameters; although we could not
check it for all the poles in the reflection factors of all the breathers
on generic excited boundary states due to the very tedious details,
it was nevertheless true in each case we examined. For solitons, one
does not expect this to be the case as in some diagrams the order
of the pole is reduced by a cancellation between pairs of diagrams
that are charge conjugate (in terms of soliton charge) analogous to
the bulk Coleman-Thun diagrams, and so only the sub-leading term survives,
which cannot be calculated so simply as above (in fact, such sub-leading
terms have never been calculated in the literature, at least to our
knowledge).

There are also known cases when a certain pole in the reflection factor
needs a combined explanation in terms of a sum of a Coleman-Thun diagram
and a boundary bound state at the same time. For an example in the
scaling Lee-Yang model cf. \cite{DTW}. Such a situation also occurs
in boundary sine-Gordon theory with Neumann boundary conditions, where
it results from the confluence of two poles and a zero in the appropriate
reflection factor as the boundary parameters are tuned to their value
at the Neumann boundary condition, and has already been treated in
detail in \cite{neumann}.

\section{Conclusions}

The perturbative approach to boundary QFT around the Neumann boundary
condition has some distinct advantages. It is very well suited for
investigation of principles of boundary QFT, and for the extension
of bulk results like LSZ reduction formulae, Landau equations and
Cutkosky rules. For many applications it may be more convenient to
start from a different boundary condition (as is usually done in the
literature), because that may reduce the number of diagrams (representing
a partial re-summation of the perturbative series around the Neumann
boundary condition) or eliminate tadpoles in case of a nontrivial
vacuum field configuration (see Appendix A). 

The main result of this paper is the derivation of the Coleman-Norton
interpretation, Cutkosky rules and their application to the boundary
bootstrap. Although the rules are derived from perturbation theory,
they are actually dependent only on some analyticity properties and
unitarity (this is especially obvious for Cutkosky rules, where the
proof explicitly relies on unitarity). Therefore one expects that
there is a non-perturbative formulation along the lines followed by
Eden et al. in the last chapter of their book \cite{ELOP}, which
would be interesting to work out in detail. Such a formulation, besides
being non-perturbative, would also give us an extension of the Cutkosky
rules to the non-physical domain and to topologies distinct from the
one considered in the proof of Section 4.

Another open issue is to formulate a complete theory of perturbative
renormalization with boundaries. In the perturbation theory around
the Neumann boundary condition, it seems relatively easy to perform
arguments similar to those in bulk theories, e.g. power counting analysis
of divergent graphs. We only considered one-loop renormalization in
boundary sine-Gordon theory in Appendix B, which was enough for a
comparison with semiclassical results in the literature. It would
be interesting to have a thorough investigation of renormalizability,
proof of locality of counter-terms, investigation of possible anomalies
of symmetries and so on. Such an investigation could provide e.g.
a proof that the tadpoles indeed restore the correct vacuum expectation
value of the field in the generic case (as shown for the toy model
in Appendix A) thereby strengthening the derivation of analytic results
from perturbation theory.

Finally let us remark that Dirichlet boundary conditions cannot be
reached by perturbation theory from the Neumann case, as this would
require taking some coupling constants to infinity. However, it is
rather trivial to extend the formalism of the paper to this case,
which only requires the inclusion of a minus sign in the reflected
propagator for fields satisfying Dirichlet boundary conditions. One
also needs to require that the corresponding asymptotic fields satisfy
Dirichlet boundary conditions, since if one insisted on an asymptotic
Neumann boundary condition, that would mean adiabatic switching off
for an infinitely strong interaction, which would be a rather odd
notion. Instead, taking the asymptotic field to satisfy Dirichlet
boundary condition enables one to carry all the formalism over to
this case. We only omitted this from the main text to maintain the
arguments and the notation as simple as possible.

\subsubsection*{Acknowledgments}

This work was partially supported by the Hungarian research funds
FKFP 0043/2001, OTKA D42209, T037674, T034299, T034512 and T043582.
GT was also supported by a Széchenyi István Fellowship, while GB and
ZB by Bolyai János Research Fellowships. We acknowledge useful comments
made by M. Mintchev.

\appendix

\makeatother \makeatletter \renewcommand\theequation{\hbox{\normalsize\Alph{section}.\arabic{equation}}} \@addtoreset{equation}{section} 

\renewcommand\thefigure{\hbox{\normalsize\Alph{section}.\arabic{figure}}} \@addtoreset{figure}{section} 

\renewcommand\thetable{\hbox{\normalsize\Alph{section}.\arabic{table}}} \@addtoreset{table}{section}

\makeatother

\section{A toy model}

In our approach, the free field has Neumann boundary conditions, and
all the boundary interactions (including their quadratic part) are
taken into account perturbatively. To illustrate how this procedure
reproduces the interacting boundary conditions, we consider a toy
model of a real scalar field $\Phi$. For simplicity, we take $1+1$
spacetime dimensions (generalization to $D+1$ is straightforward,
see at the end of the section). We take the Lagrangian to be quadratic
in the field\[
L=\int_{-\infty}^{0}dx\left[\frac{1}{2}\left(\partial_{t}\Phi\right)^{2}-\frac{1}{2}\left(\partial_{y}\Phi\right)^{2}-\frac{m^{2}}{2}\Phi^{2}\right]-\frac{\lambda}{2}\Phi(y=0,\, t)^{2}\]
so the model is exactly solvable. We remark that this toy model has
already been used in the literature to illustrate several issues of
boundary field theory \cite{mintchev}.

\subsection{Exact solution}

The equations of motion are\begin{equation}
-\partial_{t}^{2}\Phi+\partial_{y}^{2}\Phi-m^{2}\Phi=0\qquad,\qquad\left.\partial_{y}\Phi\right|_{y=0}=-\lambda\Phi(y=0,\, t)\label{eq:toy_eqm}\end{equation}
with the following solution for the modes of the field: \[
f_{\kappa}(y)=\frac{1}{\sqrt{2\pi\left(\kappa^{2}+\lambda^{2}\right)}}\left((\kappa+i\lambda)e^{i\kappa y}+(\kappa-i\lambda)e^{-i\kappa y}\right)\quad.\]
For $\lambda<0$, there also exists a normalizable mode \[
f_{B}(y)=\sqrt{2|\lambda|}e^{-\lambda y}\]
 The field has the mode expansion\[
\Phi(y,\, t)=\phi_{B}(t)f_{B}(y)+\int_{0}^{\infty}d\kappa\phi_{\kappa}(t)f_{\kappa}(y)\qquad.\]
where dynamics of the modes is\begin{eqnarray*}
\frac{d^{2}\phi_{\kappa}(t)}{dt^{2}} & = & -(\kappa^{2}+m^{2})\phi_{\kappa}(t)\\
\frac{d^{2}\phi_{B}(t)}{dt^{2}} & = & -(m^{2}-\lambda^{2})\phi_{B}(t)\end{eqnarray*}
We remark that periodicity of the boundary mode in time requires $\lambda^{2}<m^{2}$
i.e. $-m<\lambda<0$ (otherwise the vacuum $\Phi=0$ is unstable under
perturbations in the direction of the boundary mode). The Lagrangian
can be written in terms of modes as\[
L=\frac{1}{2}\int_{0}^{\infty}d\kappa\,\left(\dot{\phi}_{\kappa}(t)^{2}-(k^{2}+m^{2})\phi_{\kappa}(t)^{2}\right)+\frac{1}{2}\left(\dot{\phi}_{B}(t)^{2}-\Omega^{2}\phi_{B}(t)^{2}\right)\quad,\quad\Omega=\sqrt{m^{2}-\lambda^{2}}\]
 We can quantize the theory by introducing the conjugate momenta\[
\pi_{\kappa}(t)=\dot{\phi}_{\kappa}(t)\quad,\quad\pi_{B}(t)=\dot{\phi}_{B}(t)\]
Creation/annihilation operators can be introduced:\begin{eqnarray*}
\phi_{\kappa}(t) & = & \frac{1}{\sqrt{2\omega_{\kappa}}}\left(a(\kappa)e^{-i\omega_{\kappa}t}+a^{\dagger}(\kappa)e^{i\omega_{\kappa}t}\right)\quad,\quad\omega_{\kappa}=\sqrt{\kappa^{2}+m^{2}}\\
\phi_{B}(t) & = & \frac{1}{\sqrt{2\Omega}}\left(be^{-i\Omega t}+b^{\dagger}e^{i\Omega t}\right)\end{eqnarray*}
and satisfy\[
\left[a(\kappa),\, a^{\dagger}(\kappa')\right]=\delta(\kappa-\kappa')\quad,\quad\left[b,\, b^{\dagger}\right]=1\;.\]
For the case $\lambda>0$ (i.e. no bound states) the propagator takes
the form\begin{eqnarray*}
\left\langle 0\right|T\left(\Phi(y,\, t)\Phi(y',\, t')\right)\left|0\right\rangle  & = & \theta(t-t')\left\langle 0\right|\Phi(y,\, t)\Phi(y',\, t')\left|0\right\rangle +\theta(t'-t)\left\langle 0\right|\Phi(y',\, t')\Phi(y,\, t)\left|0\right\rangle \\
 & = & \int_{-\infty}^{\infty}\int_{-\infty}^{\infty}\frac{d\omega d\kappa}{(2\pi)^{2}}\frac{i}{\omega^{2}-\kappa^{2}-m^{2}+i\epsilon}\Bigg(e^{i\kappa(y-y')-i\omega(t-t')}\\
 &  & +\frac{\kappa-i\lambda}{\kappa+i\lambda}e^{i\kappa(y+y')-i\omega(t-t')}\Bigg)\end{eqnarray*}
from which we can read off the reflection factor\[
R(\kappa)=\frac{\kappa-i\lambda}{\kappa+i\lambda}\]
which has a bound state pole at\[
\kappa=m\sinh\vartheta=-i\lambda\]
where $\vartheta$ is the rapidity. Introducing $u=-i\vartheta$ \[
\sin u=-\frac{\lambda}{m}\]
For a bound state pole in the physical strip $0<u<\frac{\pi}{2}$
we need $-m<\lambda<0$. When this is satisfied, we get an additional
term in the propagator. This can be obtained by continuing $\lambda$
to negative values or by explicitely adding the contribution of the
bound state mode, both of which gives\begin{eqnarray}
\left\langle 0\right|T\left(\Phi(y,\, t)\Phi(y',\, t')\right)\left|0\right\rangle  & = & \theta(t-t')\left\langle 0\right|\Phi(y,\, t)\Phi(y',\, t')\left|0\right\rangle +\theta(t'-t)\left\langle 0\right|\Phi(y,\, t)\Phi(y',\, t')\left|0\right\rangle \nonumber \\
 & = & \int_{-\infty}^{\infty}\int_{-\infty}^{\infty}\frac{d\omega d\kappa}{(2\pi)^{2}}\frac{i}{\omega^{2}-\kappa^{2}-m^{2}+i\epsilon}\Bigg(e^{i\kappa(y-y')-i\omega(t-t')}\nonumber \\
 &  & +\frac{\kappa-i\lambda}{\kappa+i\lambda}e^{i\kappa(y+y')-i\omega(t-t')}\Bigg)\nonumber \\
 & + & \frac{|\lambda|}{\Omega}e^{-\lambda(y+y')}\left(\theta\left(t-t'\right)e^{-i\Omega\left(t-t'\right)}+\theta\left(t'-t\right)e^{-i\Omega\left(t'-t\right)}\right)\label{eq:toy_prop}\end{eqnarray}
The bound state term can also be written in the form\[
\int_{-\infty}^{\infty}\frac{d\omega}{2\pi}\frac{i}{\omega^{2}-\Omega^{2}+i\epsilon}e^{-i\omega(t-t')}\, f_{B}(y)f_{B}(y')\]
which means that (apart from the $y$ dependence) the boundary bound
state behaves as a free boundary field of mass $\Omega$ propagating
in $0+1$ dimensions (i.e. a harmonic oscillator) just as we postulated
for the boundary degrees of freedom in our general exposition of perturbation
theory, except that here the boundary degree of freedom arises as
a boundary bound state and not from a separate field introduced in
the Lagrangian. As a result, it is not sharply localized to the boundary,
but its contribution decreases exponentially away from $y=0$.

\subsection{Boundary perturbation theory}

The reflection factor can be expanded in the coupling $\lambda$\begin{equation}
R(\kappa)=1+2\sum_{n=1}^{\infty}\left(-\frac{i\lambda}{\kappa}\right)^{n}\label{R_exp}\end{equation}
We now proceed to show that this is correctly obtained using the perturbation
theory introduced above. We suppose that $\lambda>0$, so that no
boundary bound state exists, which would be a non-perturbative effect
which perturbation theory is not expected to reproduce.

\subsubsection{First order correction}

\begin{figure}
\begin{center}\includegraphics{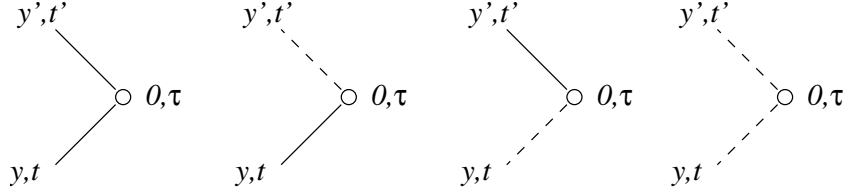}\end{center}

\caption{\label{cap:First-order-Feynman}First order Feynman graphs for the
propagator of the toy model }
\end{figure}

The interacting propagator can be written as\begin{eqnarray*}
G(y,\, t;\, y',\, t') & = & \left\langle 0\right|T\left(\Phi(y,\, t)\Phi(y',\, t')\right)\exp\left\{ -\frac{i\lambda}{2}\int_{-\infty}^{\infty}d\tau\,:\,\Phi(0,\tau)^{2}\,:\right\} \left|0\right\rangle \\
 & = & \sum_{n=0}^{\infty}\frac{1}{n!}\left(-\frac{i\lambda}{2}\right)^{n}\left\langle 0\right|T\left(\Phi(y,\, t)\Phi(y',\, t')\right)\prod_{i=1}^{n}\int_{-\infty}^{\infty}d\tau_{i}\,:\,\Phi\left(0,\tau_{i}\right)^{2}\,:\left|0\right\rangle \end{eqnarray*}
the first order correction is\begin{eqnarray*}
G^{(1)}(y,\, t;\, y',\, t') & = & -\frac{i\lambda}{2}\int_{-\infty}^{\infty}d\tau\,\left\langle 0\right|T\left(\Phi(y,\, t)\Phi(y',\, t')\right):\,\Phi\left(0,\tau\right)^{2}\,:\left|0\right\rangle \\
 & = & -i\lambda\int_{-\infty}^{\infty}d\tau\, G(y,\, t,\,0,\,\tau)G(y',\, t',\,0,\,\tau)\end{eqnarray*}
where \[
G(y,\, t,\, y',\, t')=\int_{-\infty}^{\infty}\frac{d\kappa}{2\pi}\int\frac{d\omega}{2\pi}\frac{\mathrm{e}^{-i\omega\cdot(t-t')}}{\omega^{2}-\kappa^{2}-m_{a}^{2}+i\varepsilon}\left(\mathrm{e}^{i\kappa(y-y')}+\mathrm{e}^{i\kappa(y+y')}\right)\]
is the sum of the direct and reflected propagators. This contribution
corresponds to the Feynman graphs in Fig. \ref{cap:First-order-Feynman}.We
can integrate $\tau$ out to obtain\[
\int_{-\infty}^{\infty}d\tau\, G(y,\, t-\tau)G(y',\,\tau-t')=4\int\frac{d\kappa\, d\kappa'\, d\omega}{(2\pi)^{3}}\frac{i}{\omega^{2}-\kappa^{2}-m^{2}+i\epsilon}\frac{i}{\omega^{2}-\kappa'^{2}-m^{2}+i\epsilon}e^{i\kappa y+i\kappa'y'-i\omega(t-t')}\]

The integral over $\kappa'$ can be performed using the residue theorem;
since $y'<0$, the contour must be closed in the lower half plane
$\Im m\,\kappa'<0$. Some further manipulation yields\[
4\int\frac{d\kappa\, d\omega}{(2\pi)^{2}}\frac{1}{2\kappa}\frac{i}{\omega^{2}-\kappa^{2}-m^{2}+i\epsilon}e^{i\kappa(y+y')-i\omega(t-t')}\]
which means that the reflection factor becomes, at this order\[
R(\kappa)=1-\frac{2i\lambda}{\kappa}\]
which agrees with (\ref{R_exp}).

\subsubsection{Summing up to all orders}

\begin{figure}
\begin{center}\includegraphics{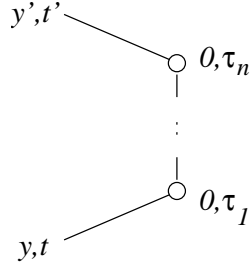}\end{center}

\caption{\label{cap:A-typical-diagram}A typical diagram of order $\lambda^{n}$.
To obtain all the diagrams, every line should be allowed to be either
a direct bulk (solid) or a reflected bulk (dashed) line, which results
in $2^{n+1}$ terms. }
\end{figure}

At the $n$th order we get the contribution\[
G^{(n)}(y,\, t;\, y',\, t')=\frac{1}{n!}\left(-\frac{i\lambda}{2}\right)^{n}\left\langle 0\right|T\left(\Phi(y,\, t)\Phi(y',\, t')\right)\prod_{i=1}^{n}\int_{-\infty}^{\infty}d\tau_{i}\,:\,\Phi\left(0,\tau_{i}\right)^{2}\,:\left|0\right\rangle \]
The relevant graphs are of the form shown in Fig. \ref{cap:A-typical-diagram}
. %
The number of contractions can be easily calculated: first one has
to decide the order in which the vertices are to be connected starting
from one end to the other: this can be done $n!$ ways. Since each
vertex has two identical legs, one has an additional factor of $2^{n}$.
Collecting all the terms we have\begin{eqnarray*}
G^{(n)}(y,\, t;\, y',\, t') & = & \frac{2^{n}n!}{n!}\left(-\frac{i\lambda}{2}\right)^{n}\int_{-\infty}^{\infty}d\tau_{1}\dots\int_{-\infty}^{\infty}d\tau_{n}\, G(y,\, t-\tau_{1})\\
 & \times & \left(\prod_{i=1}^{n-1}G\left(0,\,\tau_{i+1}-\tau_{i}\right)\right)G(y',\,\tau_{n}-t')\\
 & = & (-i\lambda)^{n}\int_{-\infty}^{\infty}d\tau_{1}\dots\int_{-\infty}^{\infty}d\tau_{n}\, G(y,\, t-\tau_{1})\left(\prod_{i=1}^{n-1}G\left(0,\,\tau_{i+1}-\tau_{i}\right)\right)G(y',\,\tau_{n}-t')\end{eqnarray*}
The propagators $G$ are the sum of the direct and reflected piece,
so the formula contains $2^{n+1}$ terms, which turn out to give identical
contributions in this case. The $\tau$ integrals can be performed
one by one, using the same method as for the first-order contribution,
with the result: \begin{equation}
G^{(n)}(y,\, t;\, y',\, t')=\int\frac{d\kappa\, d\omega}{(2\pi)^{2}}\,2^{n+1}\left(-\frac{i\lambda}{2\kappa}\right)^{n}\frac{i}{\omega^{2}-\kappa^{2}-m^{2}+i\epsilon}e^{i\kappa(y+y')-i\omega(t-t')}\label{eq:nthord_pt}\end{equation}
Summing up all the contributions, we obtain\begin{eqnarray*}
G(y,\, t;\, y',\, t') & = & \sum_{n=0}^{\infty}G^{(n)}(y,\, t;\, y',\, t')\\
 & = & \int_{-\infty}^{\infty}\int_{-\infty}^{\infty}\frac{d\omega d\kappa}{(2\pi)^{2}}\frac{i}{\omega^{2}-\kappa^{2}-m^{2}+i\epsilon}\Big(e^{i\kappa(y-y')-i\omega(t-t')}\\
 & + & R(\kappa)\, e^{i\kappa(y+y')-i\omega(t-t')}\Big)\end{eqnarray*}
where \[
R(\kappa)=1+2\sum_{n=1}^{\infty}\left(-\frac{i\lambda}{\kappa}\right)^{n}=\frac{\kappa-i\lambda}{\kappa+i\lambda}\]
which is exactly the result we expected. This means that the interacting
boundary field, built by summing up the perturbative expansion around
the free field with Neumann boundary conditions, does indeed satisfy
the correct boundary conditions (\ref{eq:toy_eqm}).

We remark that this result can be continued analytically to $\lambda<0$:
the pole at $\kappa=i\lambda$ then crosses the contour of the $\kappa$
integral, and we obtain the result (\ref{eq:toy_eqm}), which includes
the contribution from the boundary bound state. So in this model,
although the boundary bound state cannot be obtained by perturbation
theory, it can be obtained by analytically continuing the resummed
perturbative result from the regime where there is boundary bound
state. It is important to resum the perturbation series: to any finite
order the nontrivial singularity at $\kappa=-i\lambda$ in the reflection
factor is absent.

\subsubsection{Concluding remarks}

From (\ref{eq:nthord_pt}) it is apparent that there is an infrared
divergence at $\kappa=0$. This is a general feature of the perturbation
theory around the Neumann boundary condition. The perturbative expansion
is a series in $\lambda/\kappa$ and so it is convergent only for
$\kappa>\lambda$. For long wavelength modes the expansion must be
summed up and analytically continued. The physical manifestation of
this problem is that $R(\kappa=0)=-1$, while the Neumann reflection
factor is identically $+1$ for any $\kappa$. The effective strength
of the boundary interaction increases with the wavelength.

To calculate the coordinate space two-point function (\ref{eq:nthord_pt})
one needs to integrate over all wavelengths and this can only be performed
after summing up to all orders. In fact, the proper expression for
the two-point function must include an infrared regulator, which can
be removed only after summing up the leading behaviour of the function
around $\kappa=0$. As an alternative, one can perform the whole calculation
in momentum space, where every contribution is finite. 

For any finite order in perturbation theory, the amplitudes display
a singularity at $\kappa=0$, which is in fact a solution to the Landau
equations of Section 3.1 with all the bulk internal lines being on-mass-shell,
and can be accounted for using the general formalism. Summing up the
perturbation series moves this singularity to $\kappa=-i\lambda$,
and in the regime $-m<\lambda<0$ it describes a boundary excited
state. 

Finally we note that the whole analysis can be generalized to $D+1$
dimensions. The $D+1$-momentum can be written as $(\omega,\vec{k},\kappa)$
and calculations for any mode with a given value of $\vec{k}$ are
isomorphic to a calculation in the $1+1$ dimensional case, with the
particle mass $m$ replaced by $\sqrt{\vec{k}^{2}+m^{2}}$.

\subsection{Extension of the toy model: background fields}

We can extend the toy model to include a linear coupling at the boundary.
This means that there is a classical background field in the vacuum,
which poses an interesting question, since our perturbation theory
works in an expansion around the Neumann boundary condition, which
has no such fields. The question is whether with an appropriate re-summation
in the boundary coupling we can get back to the correct value of the
background field. Our Lagrangian is\[
L=\int_{-\infty}^{0}dx\left[\frac{1}{2}\left(\partial_{t}\Phi\right)^{2}-\frac{1}{2}\left(\partial_{y}\Phi\right)^{2}-\frac{m^{2}}{2}\Phi^{2}\right]-\alpha\Phi(y=0,t)-\frac{\lambda}{2}\Phi(y=0,\, t)^{2}\]
while the equations of motion take the form\begin{equation}
-\partial_{t}^{2}\Phi+\partial_{y}^{2}\Phi-m^{2}\Phi=0\qquad,\qquad\left.\partial_{y}\Phi\right|_{y=0}=-\alpha-\lambda\Phi(y=0,\, t)\label{eq:toy_eqm1}\end{equation}
which has the classical vacuum solution\begin{equation}
\phi_{vac}(x,t)=-\frac{\alpha}{m+\lambda}e^{mx}\label{eq:vacsol}\end{equation}
Let us try to compute this in perturbation theory in both $\alpha$
and $\lambda$, i.e. around the Neumann boundary condition. The diagrammatic
expression is\\

\begin{center}\begin{picture}(0,0)%
\includegraphics{vev.pstex}%
\end{picture}%
\setlength{\unitlength}{4144sp}%
\begingroup\makeatletter\ifx\SetFigFont\undefined%
\gdef\SetFigFont#1#2#3#4#5{%
  \reset@font\fontsize{#1}{#2pt}%
  \fontfamily{#3}\fontseries{#4}\fontshape{#5}%
  \selectfont}%
\fi\endgroup%
\begin{picture}(4059,942)(361,-253)
\put(361,164){\makebox(0,0)[lb]{\smash{\SetFigFont{14}{16.8}{\rmdefault}{\mddefault}{\updefault}$\left\langle\Phi(x,t)\right\rangle=$}%
}}
\end{picture}
\end{center}

where the diagrams must be understood as classes in which every bulk
line can be a direct or a reflected line (see Section 4). There is
only a first order contribution in $\alpha$, while the sum of all
orders in $\lambda$ can be easily seen to give the result of the
previous section for the two-point function, with a single $\alpha$
vertex attached to it and integrated over the boundary:\begin{eqnarray*}
\left\langle \Phi(x,t)\right\rangle  & = & -i\alpha\int_{-\infty}^{+\infty}dt\int_{-\infty}^{+\infty}\frac{d\omega}{2\pi}\int_{-\infty}^{+\infty}\frac{d\kappa}{2\pi}\frac{i}{\omega^{2}-\kappa^{2}-m^{2}+i\epsilon}e^{-i\omega t}\left[e^{i\kappa x}+R(\kappa)e^{-i\kappa x}\right]\\
R(k) & = & \frac{\kappa-i\lambda}{\kappa+i\lambda}\end{eqnarray*}
The $t$ and $\omega$ integrals can be trivially performed, and amount
to substituting $\omega=0$ in the integrand, while the $\kappa$
integral can be easily performed using the residue theorem with the
final result\[
\left\langle \Phi(x,t)\right\rangle =-\frac{\alpha}{m+\lambda}e^{mx}\]
which is fully consistent with (\ref{eq:vacsol}).

Obviously, in a general theory if there is a nontrivial vacuum solution
to the equations of motion, this means that we are going to have nontrivial
tadpole diagrams in the perturbation expansion around the free field
with Neumann boundary condition. It would be possible to eliminate
tadpoles by expanding in fluctuations around this classical background
field, but that would mean space dependent bulk couplings, which would
make analytic investigation of perturbation theory extremely difficult.
Therefore we choose to expand around the Neumann boundary condition,
since re-summation of tadpoles should give us the same result as the
background field method, as illustrated in the above example. In the
toy model it can be easily seen (just by drawing all diagrams and
using the above result for the one-point function) that all the higher
correlation functions also get the correct tadpole contributions from
perturbation theory.

\section{One loop renormalization in sine-Gordon theory}

In \cite{KP} Kormos and Palla considered the semiclassical quantization
of the two lowest energy static solutions of the boundary sine-Gordon
model:\[
V=\frac{m^{2}}{\beta^{2}}\int_{-\infty}^{0}(1-\cos\beta\Phi)\quad;\quad U=M_{0}(1-\cos\frac{\beta}{2}(\Phi-\varphi_{0}))\]
 In their work the semiclassical energy corrections are obtained by
summing up the contributions of the oscillators associated to the
linearized fluctuations around the static solutions. The appearing
standard UV divergences, which are due to the non normal ordered nature
of the Lagrangian, are canceled by counter terms: the coupling constants
are renormalized $m^{2}\to m^{2}+\delta m^{2}$ and $M_{0}\to M_{0}+\delta M_{0}$.
$\delta m^{2}$ is chosen the same as in the bulk theory (obtained
in standard perturbation theory), since due to the local nature of
the counter term it cannot depend on the presence of the boundary.
To determine $\delta M_{0}$ they impose the cancellation of logarithmic
divergences. As a result the renormalized quantities of order $\beta^{2}$
take the following form\begin{equation}
\delta m^{2}=-\frac{m^{2}\beta^{2}}{4\pi}\int_{0}^{\Lambda}\frac{dk}{\sqrt{k^{2}+m^{2}}}\quad;\quad\delta M_{0}=-\frac{M_{0}\beta^{2}}{8\pi}\int_{0}^{\Lambda}\frac{dk}{\sqrt{k^{2}+m^{2}}}\quad,\label{eq:renmm0}\end{equation}
where $\Lambda$ is the momentum cutoff. 

\begin{figure}
\begin{center}\subfigure[]{\includegraphics[%
  scale=0.5]{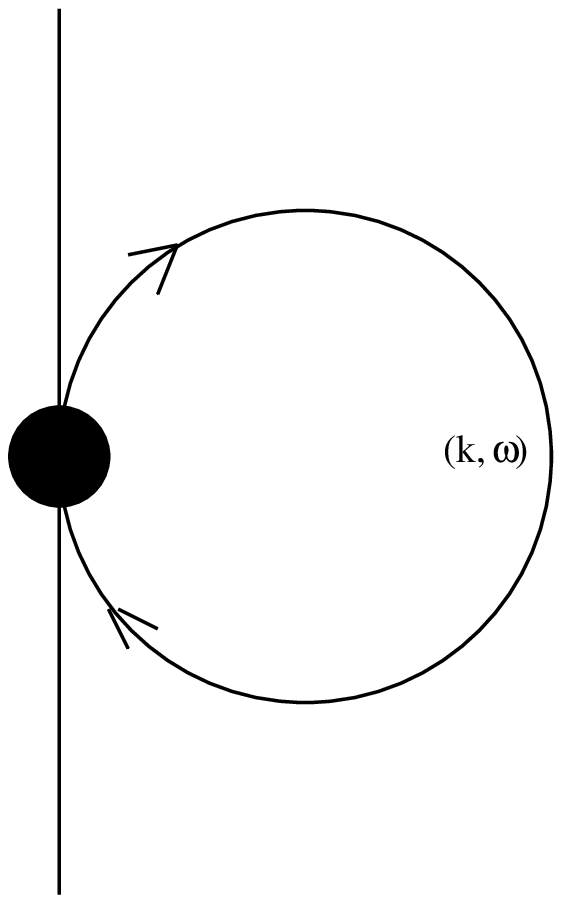}}~~~~~~~~~~~~~~~~\subfigure[]{\includegraphics[%
  scale=0.5]{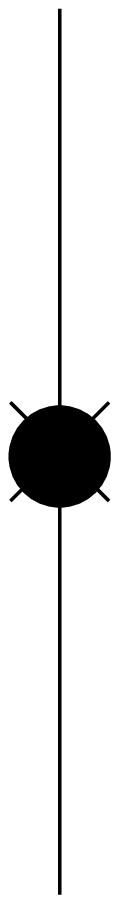}}\end{center}

\caption{\label{cap:bulk_renorm} Bulk divergence (a) and counter-term (b)
at one loop}
\end{figure}

Similar considerations lead to the same counter-terms in the analytically
continued sinh-Gordon theory \cite{CD,CTa}. The aim of this appendix
is to derive these conjectured formulae in our systematic perturbative
framework. We expand the Lagrangian to the appropriate order in $\beta^{2}$:\[
V=\frac{m^{2}}{2}\Phi^{2}-\frac{m^{2}\beta^{2}}{4!}\Phi^{4}\quad,\quad U=\frac{M_{0}\beta^{2}}{2\cdot2^{2}}\Phi^{2}-\frac{M_{0}\beta^{4}}{4!\cdot2^{4}}\Phi^{4}\]
where for simplicity we take $\varphi_{0}=0$. The divergent term
at order $m^{2}\beta^{2}$ comes from diagram \ref{cap:bulk_renorm}
(a), which is regularized by a momentum cutoff, and is canceled by
a counter-term of the form \[
V_{CT}=\frac{\delta m^{2}}{2}\Phi^{2}\]
(diagram \ref{cap:bulk_renorm} (b)). 

To compute $\delta m^{2}$ we observe that on diagram \ref{cap:bulk_renorm}
(a) the momentum $(k,\omega)$ is incoming and outgoing in the same
time, that is it does not contribute to the delta function: the integration
is unconstrained (as opposed to the case where the loop propagator
is changed to the reflected one, which results in a diagram that is
not divergent at all). As a result the delta function factor is the
same for both diagrams and from the cancellation of the vertex contribution
we have\[
2\cdot\frac{\delta m^{2}}{2}=-4\cdot3\cdot\frac{-m^{2}\beta^{2}}{4!}\int_{-\Lambda}^{\Lambda}\frac{dk}{2\pi}\int_{-\infty}^{\infty}\frac{d\omega}{2\pi}\frac{i}{\omega^{2}-k^{2}-m^{2}+i\epsilon}\]
(with the combinatorial factors given explicitly). Performing the
$\omega$ integration we recover the standard result. We have three
other divergent diagrams in this order, which can be obtained from
\ref{cap:bulk_renorm} (a) by changing any or both of its external
legs to the reflected propagator. Performing the same changes on diagram
\ref{cap:bulk_renorm} (b), however, results a diagram which removes
the required divergence.

\begin{figure}
\begin{center}\subfigure[]{\includegraphics[%
  scale=0.5]{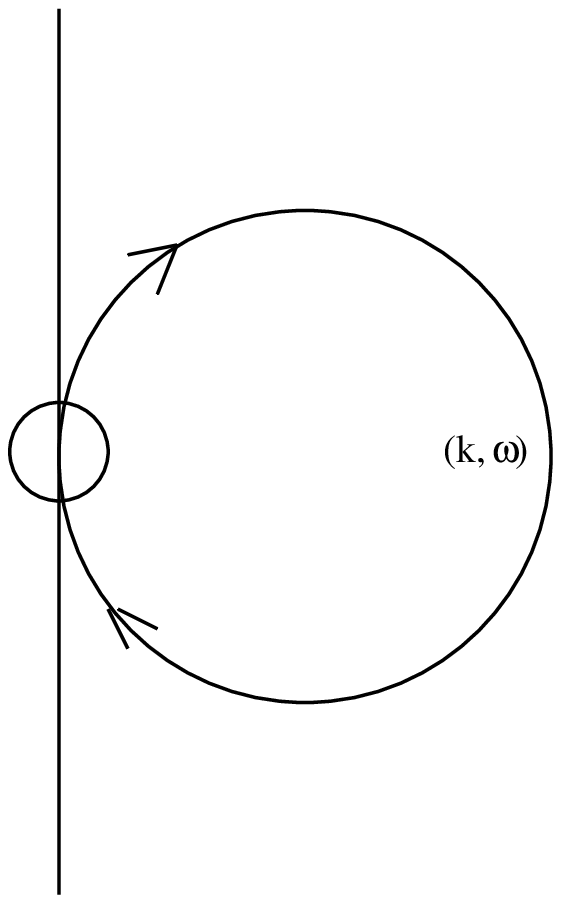}}~~~~~~~~~~~~~~~~\subfigure[]{\includegraphics[%
  scale=0.5]{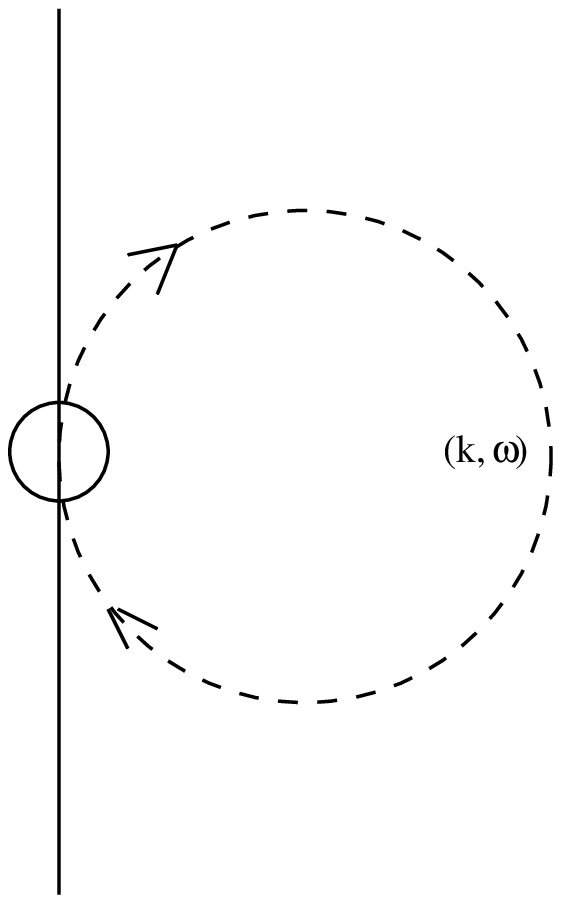}}~~~~~~~~~~~~~~~\subfigure[]{\includegraphics[%
  scale=0.5]{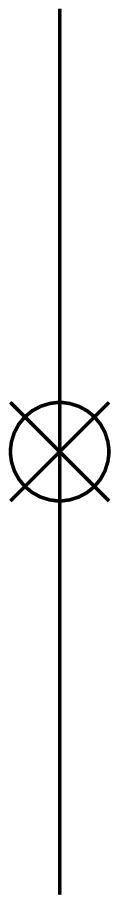}}\end{center}

\caption{\label{cap:boundary_renorm} Divergent contributions (a,b) and counter
term (c) for the boundary interaction}
\end{figure}

The divergent diagrams originating from the boundary term are of order
$M_{0}\beta^{4}$ and are canceled by a counter term of the form \[
U_{CT}=\frac{\delta M_{0}\beta^{2}}{2\cdot2^{2}}\]
The related Feynman graphs are presented on diagram \ref{cap:boundary_renorm}
(a), (b) and (c).

There are two differences compared to the bulk vertices. First of
all, $\beta$ has been replaced by $\frac{\beta}{2}$ . More importantly,
there is no momentum conservation in the boundary vertices and as
a consequence not only diagram (a) but also diagram (b) is divergent,
moreover they have the same contribution. Summing up these two terms
the counter-term acquires a factor $2$ (in addition to $\beta\rightarrow\frac{\beta}{2}$)
compared to the bulk computation and confirms the result (\ref{eq:renmm0}).

\end{document}